\documentclass[preprintnumbers, floatfix,letterpaper,aps,prd,epsfig,nofootinbib,
longbibliography,
twocolumn
]{revtex4-2}
\usepackage{bm,graphicx,dcolumn,epstopdf,epsf, latexsym,mathbbol, amssymb,amsmath, color, slashed, mathrsfs,mathcomp,simplewick}
\usepackage[center]{subfigure}
\usepackage{multirow}
\usepackage{makecell}
\usepackage{ytableau}
\usepackage{booktabs}
\usepackage[colorlinks,linkcolor=blue,citecolor=blue,urlcolor=blue]{hyperref}
\begin{document}
\allowdisplaybreaks
%%%%%%%%%%%%%%%%%%%%%%%%
 \newcommand{\bq}{\begin{equation}}
 \newcommand{\eq}{\end{equation}}
 \newcommand{\bqn}{\begin{eqnarray}}
 \newcommand{\eqn}{\end{eqnarray}}
 \newcommand{\nb}{\nonumber}
 \newcommand{\lb}{\label}
 \newcommand{\f}{\frac}
 \newcommand{\p}{\partial}
%%%%%%%%%%%%%%%%%%%%%%%%%
\newcommand{\PRL}{Phys. Rev. Lett.}
\newcommand{\PLB}{Phys. Lett. B}
\newcommand{\PRD}{Phys. Rev. D}
\newcommand{\CQG}{Class. Quantum Grav.}
\newcommand{\JCAP}{J. Cosmol. Astropart. Phys.}
\newcommand{\JHEP}{J. High. Energy. Phys.}
\newcommand{\bea}{\begin{eqnarray}}
\newcommand{\ena}{\end{eqnarray}}
\newcommand{\beqa}{\begin{eqnarray}}
\newcommand{\eeqa}{\end{eqnarray}}
\newcommand{\red}{\textcolor{red}}

\newlength\scratchlength
\newcommand\s[2]{% #1 - (up)scale-factor; #2 = content
  \settoheight\scratchlength{\mathstrut}%
  \scratchlength=\number\numexpr\number#1-1\relax\scratchlength
%  \scratchlength=\number\numexpr\number#1-1\relax\ht\strutbox
  \lower.5\scratchlength\hbox{\scalebox{1}[#1]{$#2$}}%
}

 %%%%%%%%%%%%%%%%%%%%%%%%

\title{Gravitational wave signatures of magnetized Ernst black hole}

\author{Fazlay Ahmed${}^{a}$}
\email{fazleyahmad12@gmail.com}

\author{Sanjar Shaymatov${}^{b, c}$}
\email{sanjar@astrin.uz} 

\author{Chengxun Yuan${}^{d}$}
\email{yuancx@hit.edu.cn}

\author{Salah Nasri${}^{e}$}
\email{snasri@uaeu.ac.ae, salah.nasri@cern.ch}

\affiliation{${}^{a}$Institute for Theoretical Physics \& Cosmology, Zhejiang University of Technology, Hangzhou, 310023, China\\
% ${}^{b}$ United Center for Gravitational Wave Physics (UCGWP), Zhejiang University of Technology, Hangzhou, 310023, China\\
${}^{b}$Institute of Fundamental and Applied Research,
National Research University TIIAME, Kori Niyoziy 39, Tashkent 100000, Uzbekistan\\
${}^{c}$ University of Tashkent for Applied Sciences, Str. Gavhar 1, Tashkent 100149, Uzbekistan\\
${}^{d}$ School of Physics, Harbin Institute of Technology, Harbin 150001, China\\
${}^{e}$ Department of Physics, United Arab Emirates University, Al-Ain, UAE}

\date{\today}

\begin{abstract}
We investigate gravitational wave (GW) emission from periodic timelike orbits of a test particle around a magnetized Ernst black hole and the gravitational waveforms generated by their orbital dynamics. The bound geodesics are systematically classified using the zoom-whirl representation labeled with three integers $(z,w,v)$. Gravitational waveforms are computed within a numerical framework that combines exact geodesic motion with the quadrupole approximation, which is well-suited to extreme mass-ratio inspirals (EMRIs). This analysis is particularly relevant for assessing the capability of future gravitational-wave observations to detect the effects of magnetic fields. Our results show that an intrinsic magnetic field imprints characteristic features on the GW signal, highlighting GW astronomy as a promising avenue for probing magnetized black hole spacetimes.

\end{abstract}

%\pacs{98.80.Cq, 98.80.Qc, 04.50.Kd, 04.60.Bc}

\maketitle

\section{Introduction}
\renewcommand{\theequation}{1.\arabic{equation}} \setcounter{equation}{0}

The detection of gravitational waves (GWs) by LIGO and Virgo in 2015 has revolutionized astronomy, ushering in an exciting new era of cosmic exploration. This tremendous breakthrough not only deepens our understanding of the universe but also invites us to unlock its most profound mysteries \cite{LIGOScientific:2016aoc, LIGOScientific:2016vbw, LIGOScientific:2016vlm, LIGOScientific:2016emj}. Einstein's general relativity (GR) first predicted it, which presents a distinctive observational window into the most energetic and violent cosmic events, such as binary black hole and binary neutron star mergers. Therefore, black holes have become some of the most fascinating and extensively studied objects in modern astrophysics owing to their remarkable physical properties and rich phenomenology. 

Beyond these cataclysmic phenomena, the study of particle trajectories around black holes provides a robust theoretical framework for probing the intricate dynamics of strong gravitational fields. Among these trajectories, periodic orbits are particularly significant because they play a central role in addressing fundamental challenges in astrodynamics. The analysis of periodic orbits sheds light not only on the stability of celestial systems and the complex interactions between black holes and their surrounding matter but also on generic orbital dynamics \cite{Levin:2008mq, Levin:2009sk, Misra:2010pu, Babar:2017gsg}. All generic orbits around black holes can be considered as minor deviations from periodic orbits \cite{Levin:2008mq}. The study of periodic orbits and their gravitational-wave emissions is also of particular interest because of their potential observational applications in future space-based gravitational-wave detectors. 

Black holes with stellar mass or neutron stars are often found in close orbits around supermassive black holes (SMBHs). Such binary systems are known as the extreme mass ratio inspiral (EMRI), being one of the most critical targets of future space-based gravitational detectors, such as Taiji \cite{Hu:2017mde}, Tianqin \cite{TianQin:2015yph, Gong:2021gvw}, LISA \cite{Danzmann:1997hm, Schutz:1999xj, Gair:2004iv, LISA:2017pwj, Maselli:2021men}, etc. The examination of gravitational waveform features detected by these detectors allows for a precise reconstruction of the orbital dynamics of compact objects and the surrounding gravitational field of black holes, yielding key constraints on cosmic evolution and strong-field gravity \cite{Bian:2025ifp,Ni:2024acg,Barausse14PRD,Cardoso22PRD,Zhang26EPJC,Yang24JCAP,Shabbir25,Junior24,Haroon25,Alloqulov25GW,Wang25JCAP,Lu25GW,Ahmed26GW1,Ahmed25GW2}. Given that the energy carried away by the orbital motion of the lower-mass object is an exceedingly small fraction of the total energy of the system, the time it takes for the smaller mass object to spiral around the supermassive black hole can span several years. During this process, the orbital dynamics of the smaller-mass object can be well approximated by periodic orbits \cite{Healy09PRL,Misra:2010pu,Pugliese13,Lin23,Yao23,Lin21,Tu23,Deng20,Wei19,JIANG2024,Wei25,Alloqulov26GW1,Sharipov25}.  

Also, an important aspect of astrophysical black holes is their interaction with electromagnetic fields. In GR, it has been shown that the magnetic field of a collapsing massive object decays as $t^{-1}$ \cite{Ginzburg1964,Anderson70}, implying that an isolated black hole cannot sustain an intrinsic magnetic field. Nevertheless, magnetic fields can be generated in the vicinity of black holes by external astrophysical environments, such as magnetized accretion disks \cite{Wald74} and nearby neutron stars \cite{Ginzburg1964,Rezzolla01,deFelice03}. In realistic astrophysical scenarios, black holes are expected to be embedded in external magnetic fields that may be treated as test fields satisfying $B\ll B_{max}$, such that the corresponding spacetime backreaction can be safely neglected \cite[see, e.g.,][]{Frolov10,Aliev02,Abdujabbarov10,Shaymatov14,Jamil15,Tursunov16,Hussain15,Shaymatov15,Shaymatov21pdu,Shaymatov21d,Shaymatov24PRD.110d4042S}. Typical estimates suggest magnetic field strengths of $\sim 10^{8}$ G for stellar-mass black holes and $\sim 10^{4}$ G for supermassive black holes \cite{Piotrovich10}, while similar field strengths have also been inferred near the event horizon \cite{Eatough13,Shannon13}. Independent observational analyses have reported magnetic fields ranging from 200 to $8.3 \times 10^{4}$~G at one Schwarzschild radius \cite{Baczko16}, 33.1±0.9G in the corona of the black hole binary V404 Cygni \cite{Dallilar2018}, and 1–30G around M87* from polarized emission observed by the Event Horizon Telescope \cite{EventHorizonTelescope:2021srq,EventHorizonTelescope:2021bee}. These estimates indicate that the magnetic field in the environment of astrophysical black holes remains an active subject of observational and theoretical investigation.

Since magnetic fields strongly influence the motion of charged particles through the Lorentz force, it is essential to investigate exact solutions describing the interaction between gravity and externally generated, axially symmetric magnetic fields, thereby providing a framework for testing the underlying spacetime geometry in the vicinity of black holes \cite{Aliev89}. In this context, particular attention has been devoted to static and spherically symmetric black hole solutions immersed in the Melvin magnetic universe, where the magnetic field itself contributes to the gravitational field \cite{Ernst76}. These configurations, commonly referred to as magnetized black holes \cite{Gibbons13}, have been further generalized to include rotating and electrically charged magnetized black hole spacetimes \cite{Ernst76wz,Aliev89wz,Garcia85wz}. Subsequently, a more general class of magnetized black hole solutions incorporating a global charge was developed \cite{Gibbons14wz,Astorino16wz}. These solutions have attracted considerable attention, leading to extensive investigations of their intriguing physical properties  \cite{Konoplya08a,Konoplya08b,Shaymatov21c,Shaymatov22PhRvD.106b4039S,Mirkhaydarov2026MPP,Shaymatov23GRG,2022EPJC...82..636S}.

In this work, we investigate gravitational-wave (GW) emission arising from the periodic orbital motion of a test particle in the vicinity of a magnetized Ernst black hole. The primary objective is to investigate how magnetic field effects may alter the structure of periodic orbits and imprint signatures on the resulting gravitational radiation. Our analysis indicates that an intrinsic magnetic field in the background spacetime can lead to measurable changes in the phase, amplitude, and spectral content of the emitted GWs, potentially allowing future space-based detectors such as LISA, Taiji, and TianQin to probe magnetic field effects in the strong-gravity regime \cite{Hu:2017mde,TianQin:2015yph,LISA:2017pwj}.

The paper is organized as follows: In Section~\ref{section2}, we provide a brief review of the magnetized Ernst black hole spacetime. Section~\ref{section3} is devoted to the dynamics of massive test particles in this background metric, with particular emphasis on the properties of periodic bound orbits. In Section~\ref{section4}, we compute the gravitational-wave emission from these periodic trajectories and present the corresponding characteristic strain curves and Fourier spectra. Finally, we summarize our main results and discuss their physical implications in Section~\ref{section5}.

\section{Magnetized Ernst black hole spacetime}\label{section2}
\renewcommand{\theequation}{2.\arabic{equation}} \setcounter{equation}{0}

In this section, we give a brief review of the magnetized Ernst black hole together with its magnetic field in the environment surrounding the black hole.  In this work, we focus on gravitational radiation from a particle's periodic orbits around the magnetized black hole. We begin with the line element \cite{Alonso-Bardaji:2021yls, Alonso-Bardaji:2022ear}
\begin{eqnarray}
 ds^2 &=& A^2(r,\theta)\left(-f(r)\,dt^2+ \frac{1}{f(r)}\,dr^2+r^2 d\theta^2\right) \nonumber\\&&\,+\,\frac{r^2 \sin^2\theta}{A^2(r,\theta)}d\phi^2,
\end{eqnarray}\label{metric1}
with
\begin{eqnarray}
    f(r)&=&1-\frac{2 M}{r},\\
    A(r,\theta)&=&1+B^2 r^2 \sin^2\theta\, ,
\end{eqnarray}\label{fr}
where $M$ denotes the black hole mass and $B$ is the magnetic field parameter. It is worth noting that this spacetime metric is not asymptotically flat and not spherically symmetric either. Interestingly, the event horizon is located at $r_{h}=2M$, coinciding with the Schwarzschild black hole horizon. The electromagnetic field around the magnetized Ernst black hole is written as follows 
\begin{align}\label{eq:vec-pot}
A_{\mu}dx^{\mu}=\frac{B r^2 \sin^2\theta}{2 A(r,\theta)}d\phi\, .
\end{align}
The assumption of an axially aligned magnetic field breaks the spherical symmetry of the spacetime and reduces it to axial symmetry. As a result, both the metric tensor and the electromagnetic field are invariant under rotations about the $\phi$-axis.  Hence, the magnetized Ernst spacetime is axially symmetric, being invariant under the transformation $\phi=\phi+c$, which is consistent with the symmetry of the electromagnetic field. The breaking of spherical symmetry gives rise to several interesting geometric and physical features, including the possibility of partially mimicking certain effects commonly associated with rotation. This can be demonstrated by applying the Gauss--Bonnet theorem to the horizon geometry. To this end, the metric on a constant-time hypersurface $t$ is written as \cite{2022EPJC...82..636S}
\begin{equation}
ds^2=A(r,\theta)^2\left(\frac{dr^2}{f(r)}+r^2d\theta^2\right)+\frac{r^2\sin\theta^2}{A(r,\theta)^2}d\phi^2\, .
\end{equation}
Based on the Gaussian curvature with respect to $g^{(2)}$ on $\mathcal{M}$, the Gauss-Bonnet theorem reads as 
\begin{equation}
\iint_{\mathcal{M}} K d\mathcal{A}=2 \pi \chi(\mathcal{M})\, ,
\end{equation}
where $d\mathcal{A}$ denotes the surface area element of the two-dimensional manifold, and $\chi(\mathcal{M})$ is the Euler characteristic number of the manifold. According to the Gauss--Bonnet theorem, the Euler characteristic number can be expressed in terms of the Ricci scalar of the two-dimensional surface evaluated at the event horizon, $r=r_h$, as
\begin{equation}
\mathcal{R}=\frac{2}{r^2A(r,\theta)^2}\, \mbox{~~and~~}\, {\sqrt{g^{(2)}}=r^2\sin\theta}|_{r=r_h}\, ,
\end{equation}
which leads to  
\begin{equation}
\frac{1}{4 \pi}\iint_{\mathcal{M}} \mathcal{R} \sqrt{-g^{(2)}} \,d\theta d\phi=\chi(\mathcal{M})\, .
\end{equation}
The above integral cannot be evaluated analytically because of its nontrivial dependence on $\theta$. However, for sufficiently small $B\ll1$, it can be expanded perturbatively, yielding the leading-order result
\begin{eqnarray}
\chi=2-8B^2 r_h^2/3\, .
\end{eqnarray}
This expression shows that the Euler characteristic number is smaller than its Schwarzschild value of 2, indicating that the Ernst spacetime deviates from that of a perfect sphere.

The orthonormal components of the magnetic field measured by zero-angular-momentum observers (ZAMOs), with four-velocity, are expressed as
\begin{eqnarray}
\label{b1}  B^{\hat r}
&&=-\frac{B}{A (r,\theta)}\left(1-\frac{B^2r^2\sin^2\theta}{A (r,\theta)}
\right)\cos\theta \, , \\
\label{b2}  B^{\hat\theta} &&
=\frac{BF(r)^{1/2}}{A (r,\theta)}\left(1-\frac{B^2r^2\sin^2\theta}{ A (r,\theta)}
\right)\sin\theta \, .
\end{eqnarray}
The magnetic field components are given in terms of the magnetic field parameter $B$. In the limit of \hbox{$M/r\rightarrow 0$} and
\hbox{$ A (r,\theta)\rightarrow 1$}, the magnetic field components, Eq.~(\ref{b1}) and (\ref{b2}), reduce to 
\begin{eqnarray}
 B^{\hat r} =-B\cos\theta,  ~~~ B^{\hat\theta}=B\sin\theta \, ,
\end{eqnarray}
which is similar to the homogeneous magnetic field in flat spacetime.
\begin{figure*}
\begin{tabular}{c c}
\includegraphics[scale=0.6]{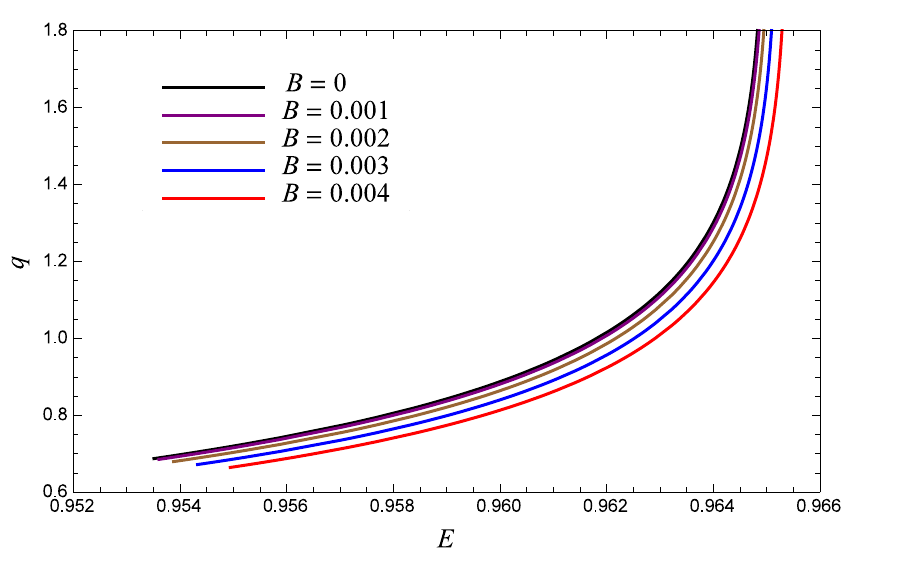}&
\includegraphics[scale=0.6]{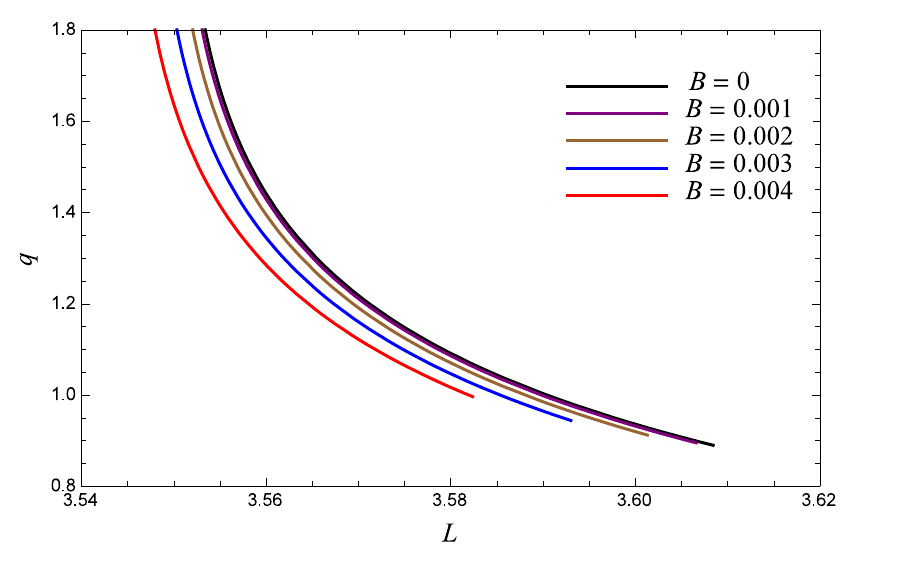}\\
\end{tabular}
\caption{The figure demonstrates the dependence of the rational number $q$ on the energy (left panel) and orbital angular momentum (right panel) for different values of the magnetic parameter $B$. Here, we set $L=3.7$  and $E=0.95$ for the left and right panels, respectively.}\label{qq1} 
\end{figure*}
\begin{figure*}
\begin{tabular}{c c c}
\includegraphics[scale=0.55]{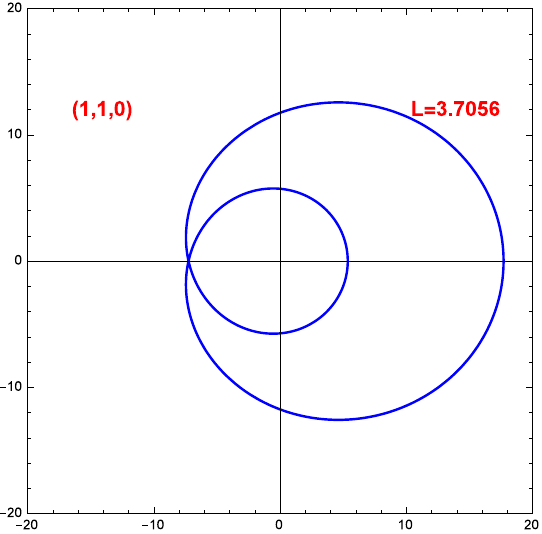}&
\includegraphics[scale=0.55]{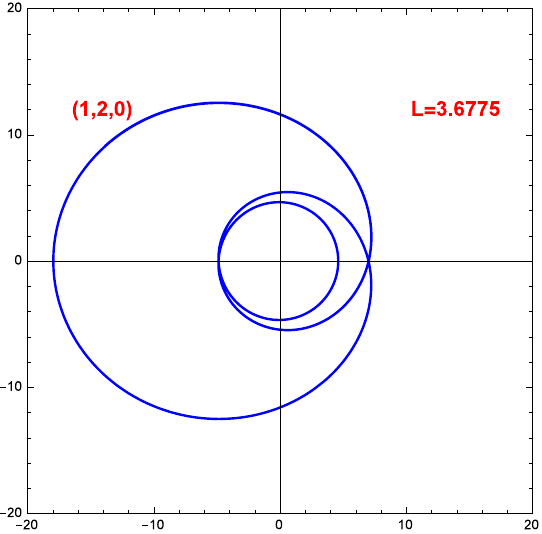}&
\includegraphics[scale=0.55]{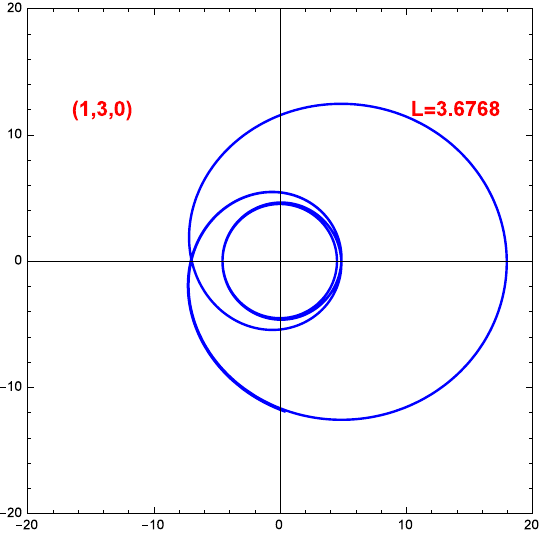}\\
\includegraphics[scale=0.55]{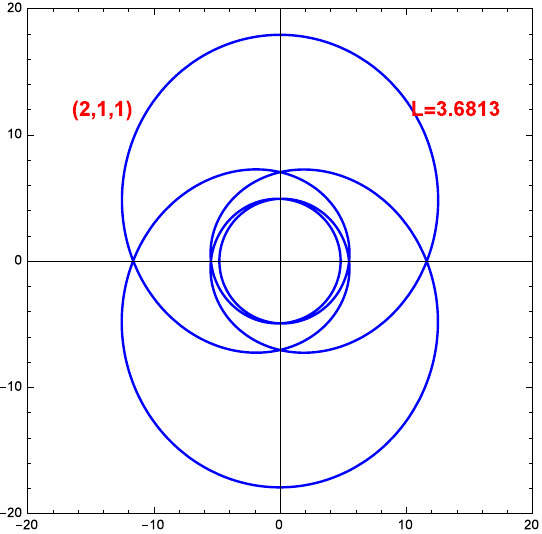}&
\includegraphics[scale=0.55]{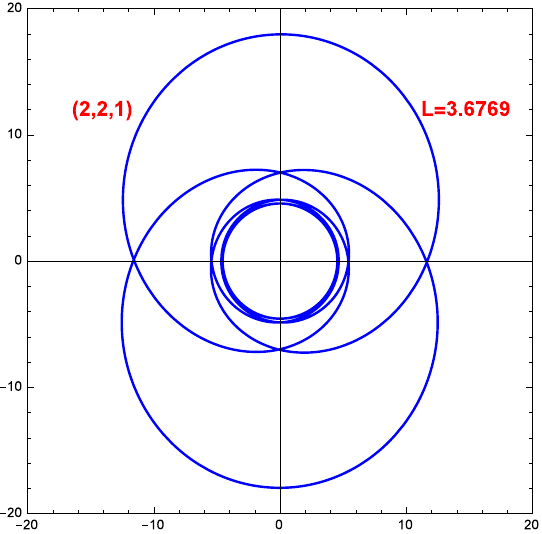}&
\includegraphics[scale=0.55]{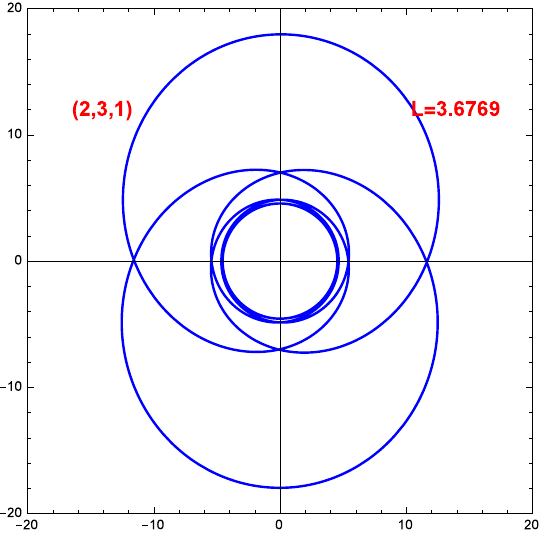}\\
\includegraphics[scale=0.55]{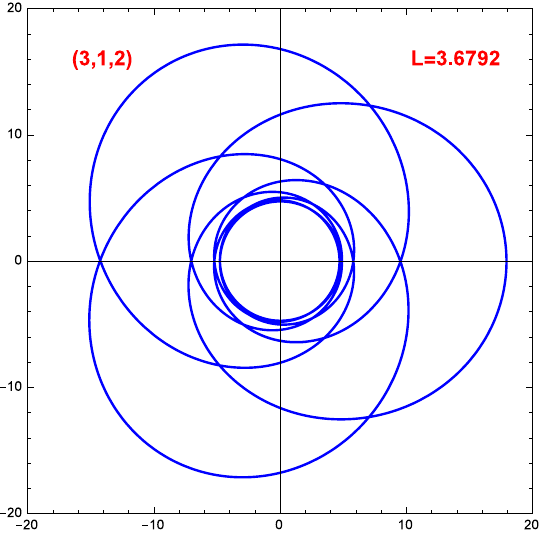}&
\includegraphics[scale=0.55]{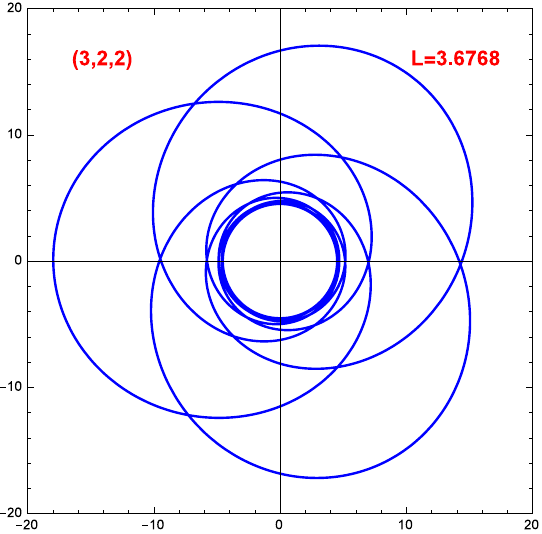}&
\includegraphics[scale=0.55]{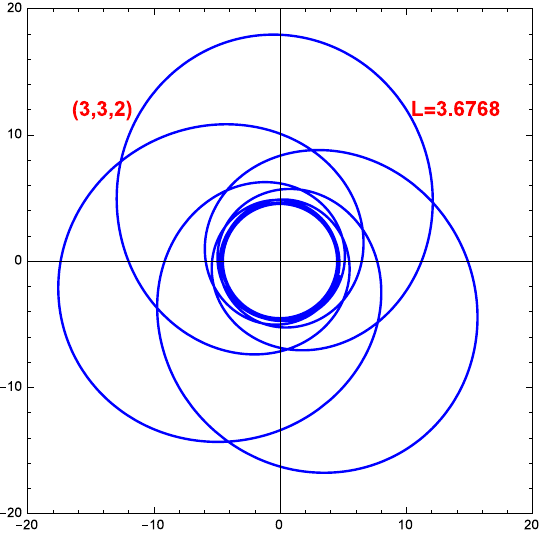}\\
\end{tabular}	
\caption{Periodic orbits around the magnetized Ernst black hole. The particle energy is fixed at $E = 0.96$. Each trajectory corresponds to a different set of zoom–whirl–vertex numbers $(z, w, v)$, illustrating the geometric complexity and structure of the bound periodic orbits.}
\label{per-orb1} 
\end{figure*}

\section{Periodic orbits}\label{section3}
\renewcommand{\theequation}{3.\arabic{equation}} \setcounter{equation}{0}

The periodic time-like orbits around the magnetized black hole are covered in this section. To determine the complex structure of bound orbits in strong gravitational fields requires a brief review of periodic orbits ~\cite{Levin:2008yp}.  Let us first consider the motion of a test particle in the background of a given black hole. The Lagrangian defines the dynamics of this particle and can be read 
\bqn
{\cal L} = \frac{1}{2} g_{\mu\nu} \frac{dx^\mu}{d\lambda} \frac{dx^\nu}{d\lambda}\, ,
\eqn
where $\tau$ denotes the proper time, which serves as the affine parameter along the world line of a timelike particle. For a massless particle, ${\cal L} = 0$, while for a massive one ${\cal L} < 0$.  

The corresponding generalized momentum $p_\mu$ is given by
\bqn
p_\mu = \frac{\partial {\cal L}}{\partial \dot{x}^\mu} = g_{\mu\nu} \dot{x}^\nu,
\eqn
which leads to the following conserved quantities for a stationary and axisymmetric spacetime:
\bqn
p_t &=& g_{tt} \dot{t} = -E,\\
p_\phi &=& g_{\phi\phi} \dot{\phi} = L,\\
p_r &=& g_{rr} \dot{r},\\
p_\theta &=& g_{\theta\theta} \dot{\theta},
\eqn
where $E$ and $L$ represent, respectively, the conserved energy and angular momentum per unit mass of the particle. A dot denotes differentiation with respect to the affine parameter $\lambda$.  

From these definitions, we obtain
\bqn\lb{dot1}
\dot{t} = -\frac{E}{g_{tt}} = \frac{E}{f(r,\theta) A^2(r)},\\
\lb{dot2}
\dot{\phi} = \frac{L}{g_{\phi\phi}} = \frac{L A^2(r)}{r^2 \sin^2\theta}.
\eqn
For timelike geodesics, the normalisation condition
\begin{equation}
g_{\mu\nu} \dot{x}^\mu \dot{x}^\nu = -1
\end{equation}
must hold. Substituting Eqs.~(\ref{dot1}) and (\ref{dot2}) into this relation yields
\bqn
g_{rr}\dot{r}^2 + g_{\theta\theta}\dot{\theta}^2 &=& -1 - g_{tt}\dot{t}^2 - g_{\phi\phi}\dot{\phi}^2 \nb\\
&=& -1 + \frac{E^2}{f(r,\theta)A^2(r)} - \frac{L^2 A^2(r)}{r^2 \sin^2\theta}\, ,\nb\\
\eqn
which defines the radial and polar motion of a test particle in the background of a magnetic black hole. The systematic analysis of these orbits provides a natural framework for classifying zoom–whirl periodic trajectories and investigating their potential observational signatures in magnetic black hole~\cite{Levin:2008yp, Levin:2009sk, Misra:2010pu}.

Here, we are interested in the motion of particles in equatorial circular orbits. For simplicity, we choose $\theta=\pi/2$ and $\dot \theta=0$. Then the above expression can be simplified into the form
\bqn\lb{rdot}
\dot r ^2 = \frac{1}{A^4(r)} (E^2 - V_{\rm eff}(r))\, ,
\eqn
where $V_{\rm eff}(r)$ denotes the effective potential and is given by
\bqn \lb{Veff}
V_{\rm eff}(r)= f(r) A^2(r) \left(1+\frac{L^2 A^2(r)}{r^2}\right)\, .
\eqn
It is evident that $V_{\rm eff}(r) \to 1$ as $r \to +\infty$, as expected for an asymptotically flat spacetime. In this case, particles with energy $E >1$ can escape to infinity. The case $E = 1$ is the critical point between bound and unbound orbits. Thus, the maximum energy that can bind a particle in orbits is $E=1$. We can obtain the trajectory of a particle by solving Eqs.~(\ref{dot1}), (\ref{dot2}), and (\ref{Veff}) to get $t$, $\phi$, and $r$ as functions of $\tau$. However, since Eq.~(\ref{rdot}) involves taking a square root, the choice of sign corresponds to whether the particle is going inward or outward, and must be specified manually before any numerical integration. A convenient equation of motion, derived from the $r-$component of the geodesic equation, can be used for numerical analysis:
\begin{eqnarray}\label{rd2}
  \ddot{r}=\frac{G'}{2 G}\dot{r}^2 -\frac{F' G}{2 F^2} E^2+\frac{G L^2}{r^3}\, ,  
\end{eqnarray}
where we define $F$ and $G$ in terms of $A(r)$ and $f(r)$ as
\begin{eqnarray}
    F=f(r)A^2(r) \;\;\;\; \mbox{and}  \; \; \; \; G=\frac{f(r)}{A^2(r)}\, ,
\end{eqnarray}
to avoid confusion and make it simple. 
The equation~\ref{rd2} is utilized for numerical integration and helps in checking the stability of circular orbits, as well as how they evolve into periodic or zoom-whirl trajectories in strong gravitational fields~\cite{Levin:2008yp, Levin:2009sk, Misra:2010pu, Chandrasekhar:1985kt}.
\begin{figure*}
\begin{tabular}{c c c}
\includegraphics[scale=0.55]{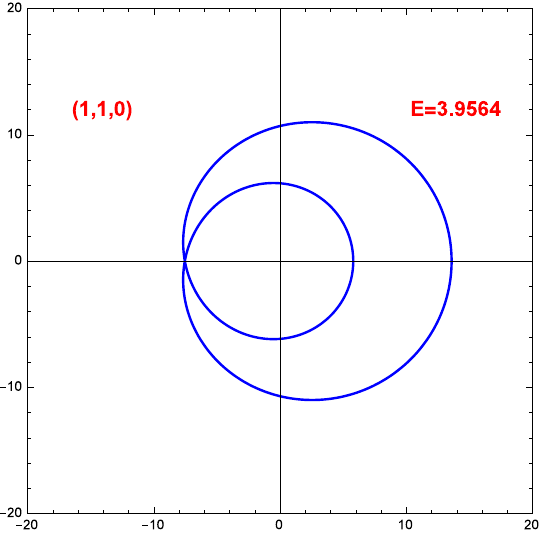}&
\includegraphics[scale=0.55]{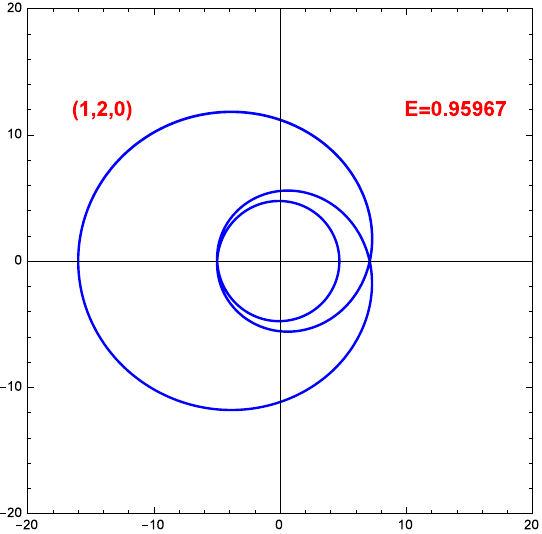}&
\includegraphics[scale=0.55]{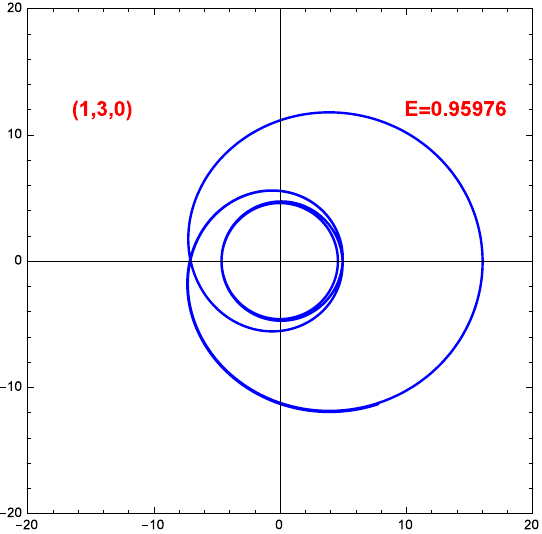}\\
\includegraphics[scale=0.55]{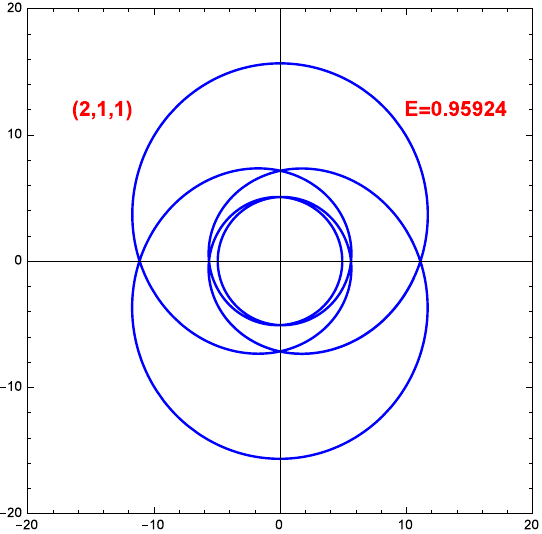}&
\includegraphics[scale=0.55]{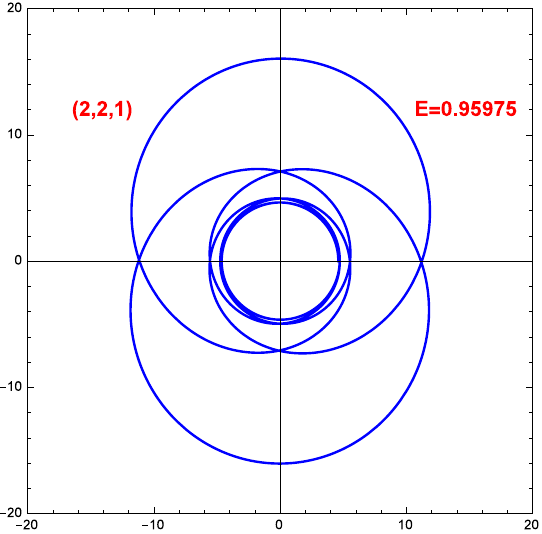}&
\includegraphics[scale=0.55]{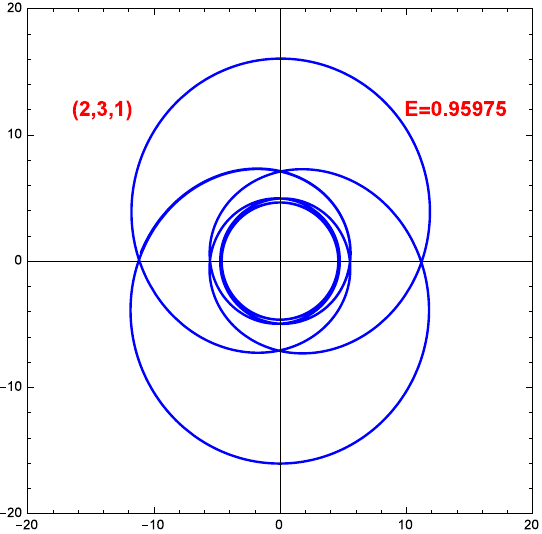}\\
\includegraphics[scale=0.55]{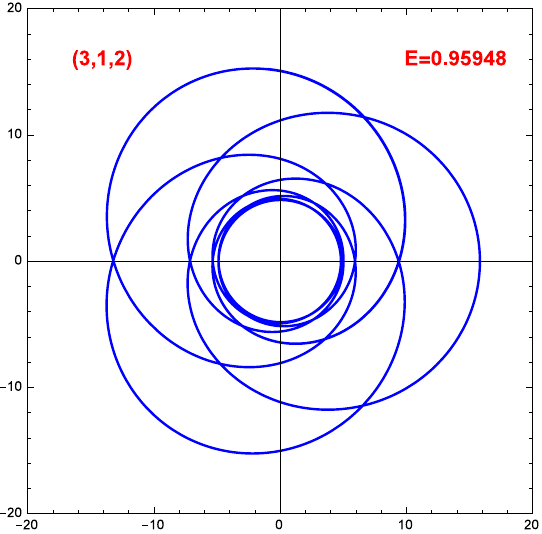}&
\includegraphics[scale=0.55]{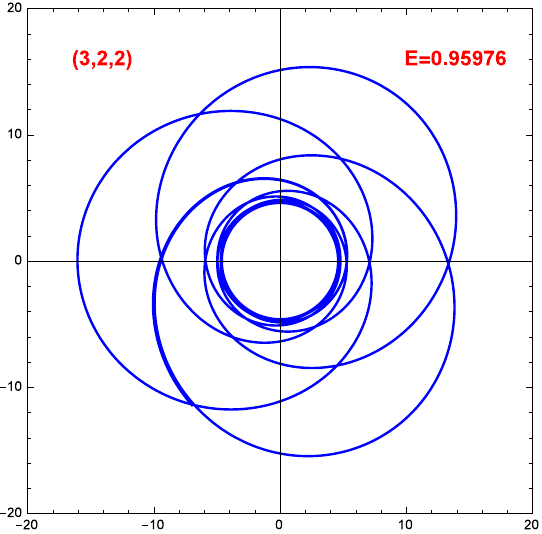}&
\includegraphics[scale=0.55]{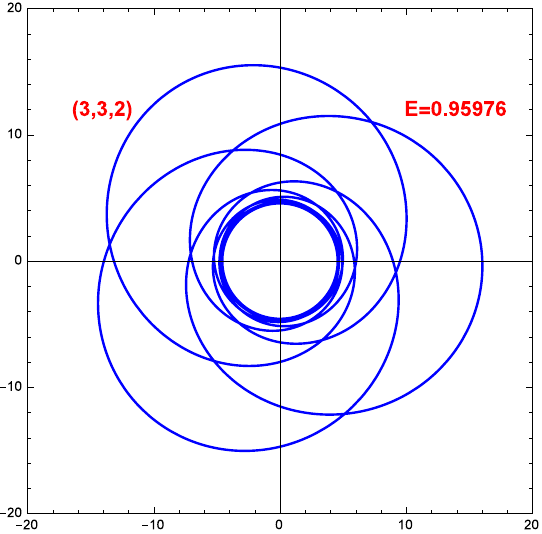}\\
\end{tabular}	
\caption{Periodic orbits for various $(z, w, v)$ combinations around the magnetized Ernst black hole. Here, we fixed the angular momentum at $L = 3.65$.}
\label{per-orb2} 
\end{figure*}

After completing the integration, a periodic orbit can be obtained for specific values of $E$ and $L$, which is a bound trajectory that returns exactly to its initial position after a fixed period. Such orbits can take various shapes, depending on the energy and angular momentum of the particle. To study them systematically, it is convenient to use a classification scheme.

Here, we utilize the approach introduced by Levin and Perez Giz~\cite{Levin:2008mq}, which classifies all periodic orbits around black holes using a representation of $(z, w, v)$, corresponding to the zoom, whirl, and vertex behavior of the trajectory. In this methodology, a periodic orbit returns to its initial position after a finite time, which implies that the ratio of the radial to azimuthal frequencies is rational. Because a rational number can approximate any irrational number, periodic orbits can effectively represent generic bound trajectories around black holes. The Levin and Perez-Giz~\cite{Levin:2008mq} method has been successfully applied to various black holes, including Schwarzschild and Kerr geometries~\cite{Levin:2009sk, Misra:2010pu, Babar:2017gsg}, and is a valuable methodology for studying the corresponding GW forms from these orbits.

According to the representations of~\cite{Levin:2008mq}, the ratio $q$ between the two frequencies $\omega_r$ and $\omega_\phi$ of oscillations in the $r$-motion and $\phi$-motion, respectively, in terms of three integers $(z, w, v)$ as
\bqn
q \equiv \frac{\omega_\phi}{\omega_r}-1 = w + \frac{v}{z}\, .
\eqn
The integers $(z, w, v)$ each have their own geometric meaning. The zoom number $z$ counts the larger circles in an orbit, while the whirl number $w$ counts the small loops near the black hole. The vertex number $v$ tells us if the particle moves through the orbit's vertices in a clockwise or anti-clockwise direction. To avoid degeneracy, $z$ and $v$ should be relatively prime~\cite{Levin:2008mq}. The parameter $q$ measures how much the orbit differs from a simple ellipse and describes its shape. This method also considers the order in which the orbital paths or segments are traced. Together, these numbers help explain the complex behavior of periodic orbits. The ratio $\frac{\omega_\phi}{\omega_r}$ equals $\Delta \phi/(2\pi)$, where $\Delta \phi = \oint d\phi$ is the total equatorial angle for one period in $r$, and this must be a multiple of $2\pi$. Using the geodesic equations for magnetized black holes, we can calculate $q$ as follows:
\bqn
q &=& \frac{1}{\pi} \int_{r_1}^{r_2} \frac{\dot \phi}{\dot r} dr -1\nb\\
&=& \frac{1}{\pi} \int_{r_2}^{r_1} \frac{L A^4(r)}{r^2\sqrt{(E^2- V_{\rm eff}(r))}}dr-1,\nonumber\\
\eqn
where $r_1$ and $r_2$ are two turning points. 

Fig.~\ref{qq1} illustrates the relationship between $q$, $E$, and $L$. The results demonstrate that energy $E$ increases monotonically with the magnetic field $B$, whereas angular momentum $L$ decreases as $B$ increases. Energy $E$ and angular momentum $L$ are restricted to a narrow limit, beyond which no other values are applicable for this analysis. These numerical values of $E$ and $L$ are essential to determine the periodic orbits.
Figures~\ref{per-orb1} and \ref{per-orb2} show the periodic orbits of black holes for different sets of integers $(z, w, v)$. The number $z$ sets how many blades the orbit has. When $z$ is larger, the blade shapes are bigger and the paths become more complex. In Figure~\ref{per-orb1}, the energy stays the same, while in Figure~\ref{per-orb2}, the angular momentum is fixed. For simplicity, we use $M=1$ in all the figures. 

\section{Gravitational Radiation in the magnetized geometry}\label{section4}
\renewcommand{\theequation}{4.\arabic{equation}} \setcounter{equation}{0}

Next, we analyze the gravitational radiation emitted by a test particle moving in periodic orbits around SMBHs, assuming that it is defined by the magnetized Ernst spacetime metric (see Eq.~\ref{metric1}). The EMRIs, consisting of a stellar-mass compact object orbiting an SMBH, are among the most promising sources for future space-based GW detectors such as LISA, Taiji, and TianQin~\cite{LISA:2017pwj, Gong:2021gvw, Hu:2017mde}. The GWs generated by these systems encode detailed information about the strong-field dynamics and the underlying spacetime geometry of the central black hole. If successful, future observations could reveal magnetic field effects in black hole spacetimes, making this research highly relevant for upcoming experimental efforts.

Exploration of gravitational wave forms from EMRIs is typically carried out using the adiabatic approximation, which implies that the inspiral timescale is much longer than the orbital period~\cite{Hughes:1999bq, Barack:2003fp}. The motion of the smaller object can be described as a series of geodesics with respect to the background metric, since its energy and angular momentum change slowly in this regime. For short-term orbital revolution~\cite{Isoyama:2021jjd}, the radiation response, or back-reaction of the emitted GWs on the particle's motion, could be neglected at leading order.
\begin{figure*}
\begin{tabular}{c c}
\includegraphics[scale=0.72]{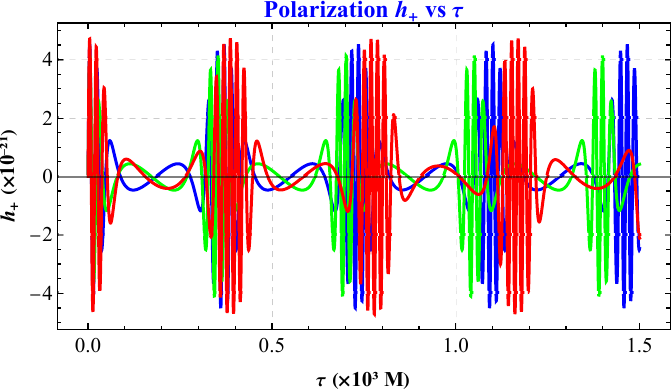}&
\includegraphics[scale=0.72]{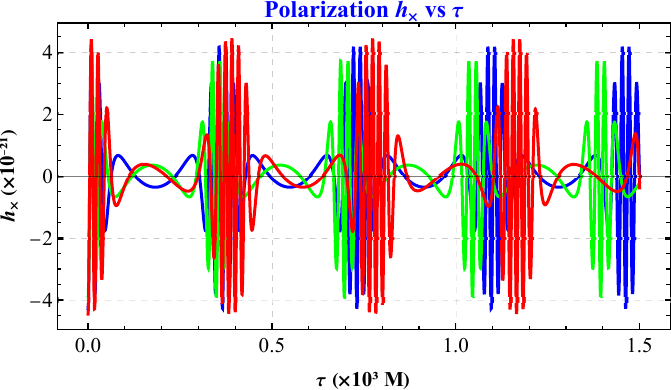}\\
\end{tabular}
\caption{GWforms (plus and cross polarizations) generated by a test particle of mass $m = 10M_{\odot}$ in periodic orbits characterized by $(z, w, v) = (1,2,0)$ (blue), $(2,1,1)$ (green), and $(3,2,2)$ (red) around a supermassive black hole of mass $M = 10^{6} M_{\odot}$. The magnetic field parameter is $B = 0.001$ and $E = 0.96$. Distinct zoom–whirl phases in the orbital motion are reflected in the modulation of the waveform amplitude and frequency.}\label{gwpolar1} 
\end{figure*}

A waveform model is employed that provides a practical method for computing gravitational waves (GWs) emitted by periodic orbits around magnetized black hole spacetimes, following the framework developed in~\cite{Babak:2006uv}, commonly known as the numerical kludge scheme. This approach involves two main steps. First, the motion of the small compact object is determined by numerically solving the geodesic equations in the black hole's background spacetime. Second, the corresponding gravitational waveform is constructed using the standard quadrupole formula for gravitational radiation. This semi-relativistic approximation has been widely adopted to model GW signals from extreme mass ratio inspirals (EMRIs) and serves as a robust tool for analyzing orbital dynamics, central black hole properties, and potential environmental effects~\cite{Barack:2003fp, Gair:2004iv, Hughes:2000ssa}.
For a metric perturbation $h_{ij}$ representing the GW and a symmetric trace-free (STF) mass quadrupole moment $I_{ij}$, the quadrupole formula takes the form
\begin{equation}
h_{ij}=\frac{1}{A}\ddot{I}_{ij},
\end{equation}
where $A=c^4 D_L/(2G)$, $G=c=1$, and $D_L$ is the luminosity distance from the source. By numerically integrating the geodesic equations, one obtains the trajectory $Z_i(t)$ of the small object in the curved spacetime of the SMBH, which is then used to compute the GW signal. For a particle of mass $m$ moving along a trajectory $Z^i(t)$, the quadrupole moment $I_{ij}$ is defined as~\cite{Thorne:1980ru}
\begin{equation}\label{lvalue}
I^{ij}=m\int d^3x\, x^i x^j\, \delta^3(x^i - Z^i(t)).
\end{equation}
\begin{figure*}
\begin{tabular}{c c}
\includegraphics[scale=0.72]{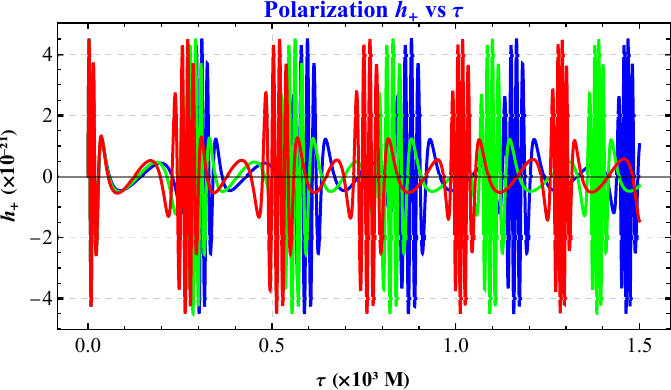}&
\includegraphics[scale=0.72]{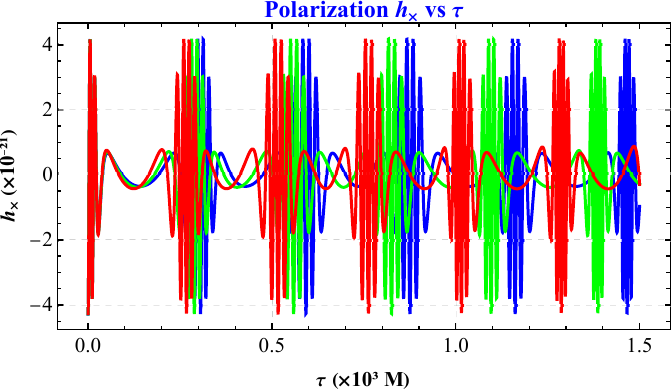}\\
\end{tabular}
\caption{Gravitational waveforms from a test object with $m=10 M_\odot$ around periodic orbits with magnetic field parameter $B=0.001$: blue, $0.002$: green, and $0.003$: red, around a supermassive black hole with mass $M=10^6 M_\odot$. The zoom-whirl value of the periodic orbit is $(1,2,0)$, and energy is fixed at $E=0.96$. The left and right panels correspond to plus and cross polarizations, respectively.} \label{gwpolar2} 
\end{figure*}

The choice of coordinate system plays a crucial role in both the computation and interpretation of gravitational waveforms. Although geodesic equations are usually solved in coordinates $(r, \theta, \phi)$, the resulting waveform is conveniently expressed in detector-adapted Cartesian coordinates $(X, Y, Z)$, which simplifies the analysis of the signal measured by a gravitational-wave detector. The transformation is given by~\cite{Babak:2006uv}
\begin{equation}\label{4.3}
x = r \sin\theta \cos\phi, \quad 
y = r \sin\theta \sin\phi, \quad 
z = r \cos\theta.
\end{equation}
This transformation enables us to project the small-object trajectory onto a Cartesian grid, which is necessary to evaluate the source multipole moments. The metric perturbations $h_{ij}$, representing the emitted GWs, are then calculated from the second time derivative of the mass quadrupole moment $I_{ij}$ as
\begin{equation}\label{4.4}
h_{ij} = \frac{m}{A}\left(a_i x_j + a_j x_i + 2 v_i v_j\right),
\end{equation}
where $v_i$ and $a_i$ denote the velocity and acceleration components of the small object, respectively, and $A = c^4 D_L / (2G)$ with $G = c = 1$. 
This formalism adheres to the conventional numerical kludge waveforms~\cite{Barack:2003fp, Gair:2004iv, Babak:2006uv, Hughes:2000ssa}, a practical, effective, and physically consistent approach to computing EMRI waveforms.

To analyze the gravitational-wave signal recorded by a detector, it is convenient to introduce a detector-adapted Cartesian coordinate system $(X, Y, Z)$, centred on the black hole and oriented with respect to the source frame $(x, y, z)$ by the inclination angle $\iota$ and the longitude of pericentre $\zeta$~\cite{Babak:2006uv, Barack:2003fp, Gair:2004iv}. This transformation facilitates the projection of the waveform onto the detector frame, enabling the computation of the observable GW polarizations.
 The unit vectors of the detector frame in the $(x, y, z)$ coordinates are:
\begin{eqnarray}
\hat{e}_X &=& (\cos\zeta, -\sin\zeta, 0),\\
\hat{e}_Y &=& (\sin\iota \sin\zeta, \cos\iota \cos\zeta, -\sin\iota),\\
\hat{e}_Z &=& (\sin\iota \sin\zeta, -\sin\iota \cos\zeta, \cos\iota),
\end{eqnarray}
The GW polarizations $h_+$ and $h_\times$ are then obtained by projecting $h_{ij}$, Eq. (\ref{4.4}), onto the detector frame
\begin{eqnarray}\label{4.5}
h_+&=\frac{1}{2}\big(e_X^i e_X^j-e_Y^ie_Y^j\big)h_{ij},\\
h_{\times}&=\frac{1}{2}\big(e_X^i e_Y^j-e_Y^ie_X^j\big)h_{ij},
\end{eqnarray}
These polarizations can be expressed in terms of components $h_{\zeta\zeta}$, $h_{\iota\iota}$, and $h_{\iota\zeta}$, which are defined in the detector frame as combinations of components $h_{ij}$ as
\begin{eqnarray}\label{4.64}
h_+&=&\frac{1}{2}\big(h_{\zeta\zeta}-h_{\iota\iota}\big),\\\label{4.6}
h_{\times}&=&h_{\iota\zeta}\, ,
\end{eqnarray}
where the components are \cite{Babak:2006uv}
\begin{eqnarray}\label{4.7}
h_{\zeta\zeta}&=&h_{xx}\cos^2\zeta-h_{xy}\sin{2 \zeta}+h_{yy}\sin^2\zeta,\\
h_{\iota\iota}&=& \cos^2\iota\big[h_{xx}\sin^2\zeta + h_{xy}\sin 2 \zeta + h_{yy}\cos^2 \zeta\big] \nonumber \\
&& +h_{zz} \sin^2\iota - \sin{2 \iota}\big[h_{xz \sin\zeta}+h_{yz}\cos\zeta\big],\\
h_{\iota\zeta}&=& \frac{1}{2}\cos\iota\big[h_{xx} \sin {2 \zeta}+ 2 h_{xy}\cos{2 \zeta}- h_{yy}\sin{2 \zeta}\big]\nonumber\\
&&+\sin\iota\big[h_{yz} \sin\zeta-h_{xx} \cos\zeta\big].
\end{eqnarray}
\begin{figure*}
\begin{tabular}{c c}
\includegraphics[scale=0.95]{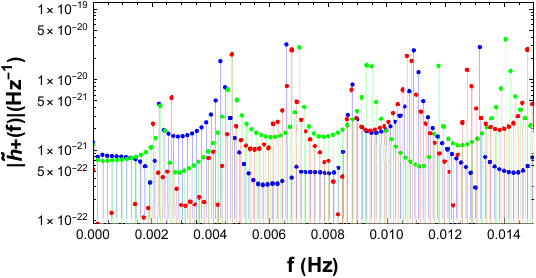}& 
\includegraphics[scale=0.95]{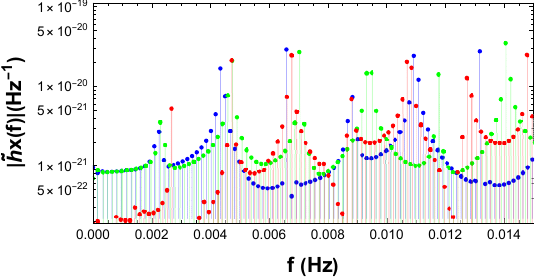}\\
\end{tabular}
\caption{Fourier spectra $|\tilde{h}_{+,\times}(f)|$ corresponding to the time-domain waveforms shown in Fig.~\ref{gwpolar1} for $B = 0.001$. The spectral peaks correspond to characteristic frequencies of the zoom–whirl orbits, showing distinct harmonic structures related to the orbital parameters $(z, w, v)$.}\label{freq-spect1} 
\end{figure*}
\begin{figure*}
\begin{tabular}{c c}
\includegraphics[scale=0.95]{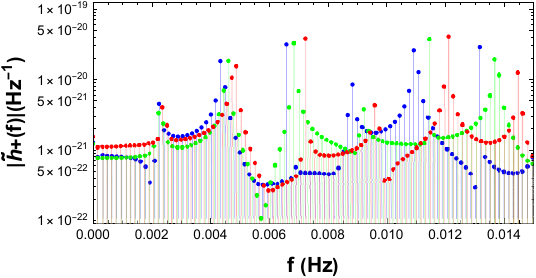}& 
\includegraphics[scale=0.95]{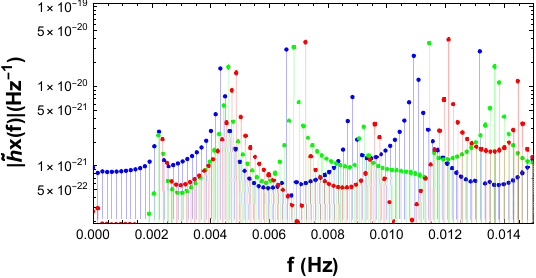}\\
\end{tabular}
\caption{Fourier spectra $|\tilde{h}_{+,\times}(f)|$ for the gravitational waveforms in Fig.~\ref{gwpolar2} with zoom-whirl value $(1,2,0))$. Increasing the magnetic field parameter $B$ shifts the spectral peaks and enhances the high-frequency components, indicating stronger gravitational radiation and modified orbital dynamics.}\label{freq-spect2} 
\end{figure*}

To examine the influence of the magnetic field on gravitational waveforms generated by different periodic orbits in an EMRI system, we consider a compact object of mass $m = 10\, M_\odot$ orbiting SMBH of mass $M = 10^6\, M_\odot$. For simplicity, the inclination angle $\iota$ and the longitude of pericentre $\zeta$ are fixed at $\pi/4$, and a luminosity distance of $D_L = 200$~Mpc is adopted for the computation of the GW polarizations. 

The resulting GW forms, described by the two independent components $h_+$ and $h_\times$, show a clear alternating pattern. When the orbit stretches outward in a highly eccentric way (the zoom phases), the waveform amplitude stays relatively low. These periods are followed by short, strong bursts of radiation that happen during the nearly circular parts of the orbit (the whirl phases). The number of low-amplitude periods matches the number of zoom segments, and the number of strong bursts matches the number of whirls in the orbit. The numerical results from Eqs.~(\ref{4.64}) and~(\ref{4.4}) appear in Figs.~\ref{gwpolar1} and~\ref{gwpolar2}, which clearly show the separate "zoom" and "whirl" features of the GW signal from periodic orbits in EMRIs. These features reflect the orbital dynamics of the small object over one full cycle~\cite{Levin:2008mq, Barack:2003fp, Babak:2006uv, Gair:2004iv}.

Figure~\ref{gwpolar1} presents gravitational waveforms corresponding to $(z,w,v)=(1,2,0)$, $(2,1,1)$, and $(3,2,2)$. The analysis defines a strong correlation between the gravitational waveforms and the orbital motion of the small object. Each orbit shows distinct "zoom" and "whirl" phases in the waveform, which correspond to analogous behaviors in the object's trajectory. Furthermore, orbits with higher zoom numbers $z$ generate waveforms with more intricate substructures, indicating an increased number of ``leaves" in the complete periodic orbit.

The magnetic field $B$ has a clear effect on the gravitational waveform produced by a small object in a periodic orbit. Figure~\ref{gwpolar2} shows how changing $B$ alters the waveform. As $B$ increases, we observe significant changes in amplitude and a noticeable phase shift, which highlights how the magnetic field influences both the orbit and the emitted radiation.

The GWs emitted by a test particle in periodic motion around an SMBH in the magnetized spacetime can be further analyzed through their frequency spectra $\vert \tilde{h}_{+,\times}(f)\vert$ and characteristic strain $h_c(f)$, defined as
\begin{eqnarray}\label{ch}
h_c(f)=2f\left(\vert \tilde{h}_+(f)\vert^2+\vert \tilde{h}_\times(f)\vert^2\right)^{1/2}.
\end{eqnarray}
We use a discrete Fourier transform (DFT) on the time-domain GWforms to get the frequency spectra, which lets us study the signal’s frequency content in detail. This process shows how the particle’s periodic orbital motion changes the structure of the emitted GWs (see Figs.~\ref{freq-spect1} and~\ref{freq-spect2}). Most of the main frequencies are in the millihertz range, so they are especially important for space-based detectors like LISA, Taiji, and Tianqin~\cite{LISA:2017pwj, Barack:2003fp, Gair:2004iv, Babak:2006uv}, which are built to detect low-frequency GWs from EMRIs. 
The characteristic spectra for different periodic orbits, labeled by the triplet, are illustrated in Fig.~\ref{freq-spect1}, while for different values of $B$ are shown in Fig.~\ref{freq-spect2}.
\begin{figure*}
\begin{tabular}{c c}
\includegraphics[scale=0.65]{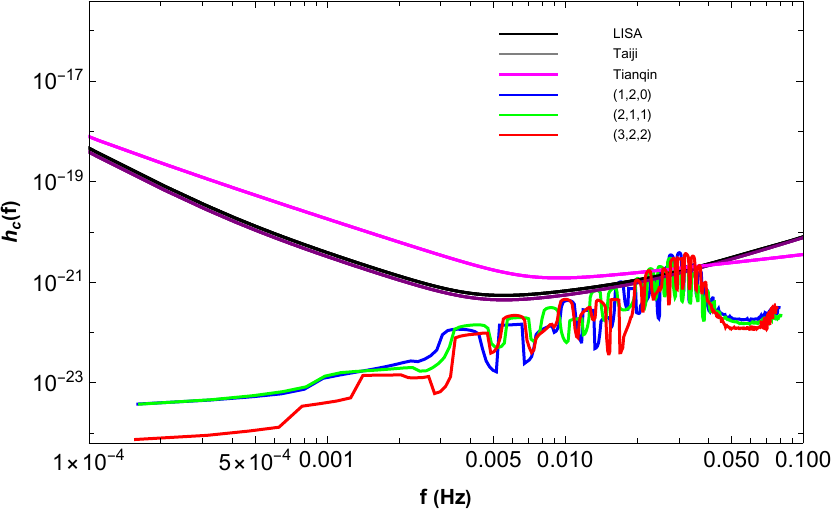}& 
\includegraphics[scale=0.65]{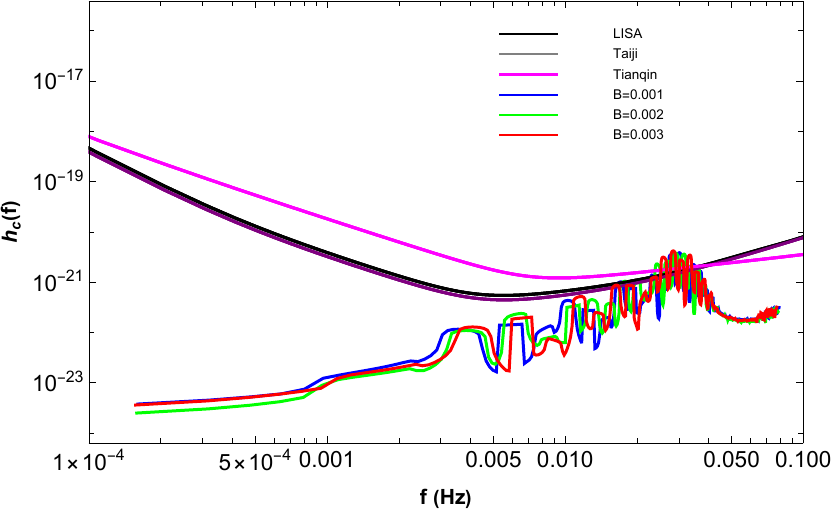}\\
\end{tabular}
\caption{Characteristics strain of gravitational waveforms of periodic orbits in Figs.~\ref{gwpolar1} (left) and \ref{gwpolar2} (right). The black, gray, and magenta curves correspond to the sensitivities of the LISA, Taiji, and Tianqin detectors. Portions of the spectra lie above the sensitivity bands, suggesting that the magnetic field effect could be detectable in future space-based GW observations.}\label{strain} 
\end{figure*}

To make the plots of the characteristic strain from Eq.~(\ref{ch}) clearer, we smooth the numerically generated $h_c(f)$ by taking a running average over 30 frequency bins. Using a larger window would reduce more numerical noise but might hide important spectral details. As shown in Fig.~\ref{strain}, some parts of the characteristic strain for different orbital setups $(z, w, v)$ and magnetic field parameter values $B$ are above the sensitivity curves of the LISA, Taiji, and Tianqin detectors. This means the related gravitational waves, which show unique zoom–whirl features from the spacetime described above, are within the range these future space-based detectors can observe~\cite{LISA:2017pwj, Gair:2017ynp, Babak:2017tow}. Detecting these signals would give us a valuable chance to study the spacetime around SMBHs and to look for possible magnetic field effects through detailed gravitational wave observations.

\section{Discussions and Conclusions}\label{section5}

GWs emitted by compact objects orbiting black holes serve as a powerful probe of strong-field gravity and the structure of spacetime. EMRIs are anticipated to be among the most informative sources for space-based detectors such as LISA, Taiji, and Tianqin, as they provide sensitive information about the background geometry through the orbital motion of the compact body. In light of this, the present study investigates the dynamics of particles around the magnetized Ernst black hole and the resulting GW signatures. Understanding these effects is essential for assessing whether future GW observations can detect signatures of magnetic fields or deviations from general relativity in the strong-field regime.

This study investigates periodic orbits and their corresponding waveforms within the constraints of geometry, including the influence of the magnetic field. The geodesics in the background spacetime were solved analytically. Subsequently, an exceptional representation \cite{Levin:2008mq} was employed to distinguish different types of periodic orbits around the magnetized black hole. In this framework, each periodic orbit is characterized by the parameters $(z, w, v)$. These results may provide a means to distinguish between magnetized black holes and the Schwarzschild black hole. An EMRI system was analyzed, consisting of a test object with mass $m = 10 M_\odot$ following periodic orbits around an SMBH with mass $M = 10^6 M_\odot$. The well-known numerical kludge scheme was used to investigate the resulting gravitational waveforms, with the system stationed at a luminosity distance of $D_L = 200$ Mpc from the detector, an inclination angle of $\iota = \pi/4$, and a longitude of pericenter $\zeta = \pi/4$. A clear correlation was demonstrated between the gravitational waveforms emitted by an object orbiting an SMBH and the object's zoom-whirl orbital behavior. Higher zoom-whirl numbers correspond to more complex waveform substructures. 

Furthermore, the presence of $B$ was shown to significantly affect these waveforms. To evaluate the detectability of gravitational waves (GWs) from EMRIs with periodic orbits, their time-domain waveforms were analyzed using discrete Fourier transforms to extract the frequency spectra. The results indicate that the frequencies of these GWs generally fall within the sensitivity range of space-based detectors. From the spectra, the characteristic strains were determined, and for certain combinations of $(z,w,v)$, the strains exceed the sensitivity thresholds of LISA, Taiji, and Tianqin. These findings suggest that space-based gravitational-wave observatories could detect signals from EMRIs with periodic orbits, thereby providing a promising approach for exploring SMBHs in spacetimes characterized by a magnetic field parameter. In summary, the presence of the magnetic field parameter plays a critical role in shaping GW signals and offers promising opportunities for future observations to probe their influence in strong gravitational fields.

%\section*{Acknowledgements}

\appendix

%\bibliography{sample}

\bibliographystyle{apsrev4-1}
\bibliography{main}

\begin{thebibliography}{92}%
\makeatletter
\providecommand \@ifxundefined [1]{%
 \@ifx{#1\undefined}
}%
\providecommand \@ifnum [1]{%
 \ifnum #1\expandafter \@firstoftwo
 \else \expandafter \@secondoftwo
 \fi
}%
\providecommand \@ifx [1]{%
 \ifx #1\expandafter \@firstoftwo
 \else \expandafter \@secondoftwo
 \fi
}%
\providecommand \natexlab [1]{#1}%
\providecommand \enquote  [1]{``#1''}%
\providecommand \bibnamefont  [1]{#1}%
\providecommand \bibfnamefont [1]{#1}%
\providecommand \citenamefont [1]{#1}%
\providecommand \href@noop [0]{\@secondoftwo}%
\providecommand \href [0]{\begingroup \@sanitize@url \@href}%
\providecommand \@href[1]{\@@startlink{#1}\@@href}%
\providecommand \@@href[1]{\endgroup#1\@@endlink}%
\providecommand \@sanitize@url [0]{\catcode `\\12\catcode `\$12\catcode
  `\&12\catcode `\#12\catcode `\^12\catcode `\_12\catcode `\%12\relax}%
\providecommand \@@startlink[1]{}%
\providecommand \@@endlink[0]{}%
\providecommand \url  [0]{\begingroup\@sanitize@url \@url }%
\providecommand \@url [1]{\endgroup\@href {#1}{\urlprefix }}%
\providecommand \urlprefix  [0]{URL }%
\providecommand \Eprint [0]{\href }%
\providecommand \doibase [0]{http://dx.doi.org/}%
\providecommand \selectlanguage [0]{\@gobble}%
\providecommand \bibinfo  [0]{\@secondoftwo}%
\providecommand \bibfield  [0]{\@secondoftwo}%
\providecommand \translation [1]{[#1]}%
\providecommand \BibitemOpen [0]{}%
\providecommand \bibitemStop [0]{}%
\providecommand \bibitemNoStop [0]{.\EOS\space}%
\providecommand \EOS [0]{\spacefactor3000\relax}%
\providecommand \BibitemShut  [1]{\csname bibitem#1\endcsname}%
\let\auto@bib@innerbib\@empty
%</preamble>
\bibitem [{\citenamefont {Abbott}\ \emph
  {et~al.}(2016{\natexlab{a}})\citenamefont {Abbott} \emph
  {et~al.}}]{LIGOScientific:2016aoc}%
  \BibitemOpen
  \bibfield  {author} {\bibinfo {author} {\bibfnamefont {B.~P.}\ \bibnamefont
  {Abbott}} \emph {et~al.} (\bibinfo {collaboration} {LIGO Scientific,
  Virgo}),\ }\href {\doibase 10.1103/PhysRevLett.116.061102} {\bibfield
  {journal} {\bibinfo  {journal} {Phys. Rev. Lett.}\ }\textbf {\bibinfo
  {volume} {116}},\ \bibinfo {pages} {061102} (\bibinfo {year}
  {2016}{\natexlab{a}})},\ \Eprint {http://arxiv.org/abs/1602.03837}
  {arXiv:1602.03837 [gr-qc]} \BibitemShut {NoStop}%
\bibitem [{\citenamefont {Abbott}\ \emph
  {et~al.}(2016{\natexlab{b}})\citenamefont {Abbott} \emph
  {et~al.}}]{LIGOScientific:2016vbw}%
  \BibitemOpen
  \bibfield  {author} {\bibinfo {author} {\bibfnamefont {B.~P.}\ \bibnamefont
  {Abbott}} \emph {et~al.} (\bibinfo {collaboration} {LIGO Scientific,
  Virgo}),\ }\href {\doibase 10.1103/PhysRevD.93.122003} {\bibfield  {journal}
  {\bibinfo  {journal} {Phys. Rev. D}\ }\textbf {\bibinfo {volume} {93}},\
  \bibinfo {pages} {122003} (\bibinfo {year} {2016}{\natexlab{b}})},\ \Eprint
  {http://arxiv.org/abs/1602.03839} {arXiv:1602.03839 [gr-qc]} \BibitemShut
  {NoStop}%
\bibitem [{\citenamefont {Abbott}\ \emph
  {et~al.}(2016{\natexlab{c}})\citenamefont {Abbott} \emph
  {et~al.}}]{LIGOScientific:2016vlm}%
  \BibitemOpen
  \bibfield  {author} {\bibinfo {author} {\bibfnamefont {B.~P.}\ \bibnamefont
  {Abbott}} \emph {et~al.} (\bibinfo {collaboration} {LIGO Scientific,
  Virgo}),\ }\href {\doibase 10.1103/PhysRevLett.116.241102} {\bibfield
  {journal} {\bibinfo  {journal} {Phys. Rev. Lett.}\ }\textbf {\bibinfo
  {volume} {116}},\ \bibinfo {pages} {241102} (\bibinfo {year}
  {2016}{\natexlab{c}})},\ \Eprint {http://arxiv.org/abs/1602.03840}
  {arXiv:1602.03840 [gr-qc]} \BibitemShut {NoStop}%
\bibitem [{\citenamefont {Abbott}\ \emph
  {et~al.}(2016{\natexlab{d}})\citenamefont {Abbott} \emph
  {et~al.}}]{LIGOScientific:2016emj}%
  \BibitemOpen
  \bibfield  {author} {\bibinfo {author} {\bibfnamefont {B.~P.}\ \bibnamefont
  {Abbott}} \emph {et~al.} (\bibinfo {collaboration} {LIGO Scientific,
  Virgo}),\ }\href {\doibase 10.1103/PhysRevLett.116.131103} {\bibfield
  {journal} {\bibinfo  {journal} {Phys. Rev. Lett.}\ }\textbf {\bibinfo
  {volume} {116}},\ \bibinfo {pages} {131103} (\bibinfo {year}
  {2016}{\natexlab{d}})},\ \Eprint {http://arxiv.org/abs/1602.03838}
  {arXiv:1602.03838 [gr-qc]} \BibitemShut {NoStop}%
\bibitem [{\citenamefont {Levin}\ and\ \citenamefont
  {Perez-Giz}(2008)}]{Levin:2008mq}%
  \BibitemOpen
  \bibfield  {author} {\bibinfo {author} {\bibfnamefont {J.}~\bibnamefont
  {Levin}}\ and\ \bibinfo {author} {\bibfnamefont {G.}~\bibnamefont
  {Perez-Giz}},\ }\href {\doibase 10.1103/PhysRevD.77.103005} {\bibfield
  {journal} {\bibinfo  {journal} {Phys. Rev. D}\ }\textbf {\bibinfo {volume}
  {77}},\ \bibinfo {pages} {103005} (\bibinfo {year} {2008})},\ \Eprint
  {http://arxiv.org/abs/0802.0459} {arXiv:0802.0459 [gr-qc]} \BibitemShut
  {NoStop}%
\bibitem [{\citenamefont {Levin}(2009)}]{Levin:2009sk}%
  \BibitemOpen
  \bibfield  {author} {\bibinfo {author} {\bibfnamefont {J.}~\bibnamefont
  {Levin}},\ }\href {\doibase 10.1088/0264-9381/26/23/235010} {\bibfield
  {journal} {\bibinfo  {journal} {Class. Quant. Grav.}\ }\textbf {\bibinfo
  {volume} {26}},\ \bibinfo {pages} {235010} (\bibinfo {year} {2009})},\
  \Eprint {http://arxiv.org/abs/0907.5195} {arXiv:0907.5195 [gr-qc]}
  \BibitemShut {NoStop}%
\bibitem [{\citenamefont {Misra}\ and\ \citenamefont
  {Levin}(2010)}]{Misra:2010pu}%
  \BibitemOpen
  \bibfield  {author} {\bibinfo {author} {\bibfnamefont {V.}~\bibnamefont
  {Misra}}\ and\ \bibinfo {author} {\bibfnamefont {J.}~\bibnamefont {Levin}},\
  }\href {\doibase 10.1103/PhysRevD.82.083001} {\bibfield  {journal} {\bibinfo
  {journal} {Phys. Rev. D}\ }\textbf {\bibinfo {volume} {82}},\ \bibinfo
  {pages} {083001} (\bibinfo {year} {2010})},\ \Eprint
  {http://arxiv.org/abs/1007.2699} {arXiv:1007.2699 [gr-qc]} \BibitemShut
  {NoStop}%
\bibitem [{\citenamefont {Babar}\ \emph {et~al.}(2017)\citenamefont {Babar},
  \citenamefont {Babar},\ and\ \citenamefont {Lim}}]{Babar:2017gsg}%
  \BibitemOpen
  \bibfield  {author} {\bibinfo {author} {\bibfnamefont {G.~Z.}\ \bibnamefont
  {Babar}}, \bibinfo {author} {\bibfnamefont {A.~Z.}\ \bibnamefont {Babar}}, \
  and\ \bibinfo {author} {\bibfnamefont {Y.-K.}\ \bibnamefont {Lim}},\ }\href
  {\doibase 10.1103/PhysRevD.96.084052} {\bibfield  {journal} {\bibinfo
  {journal} {Phys. Rev. D}\ }\textbf {\bibinfo {volume} {96}},\ \bibinfo
  {pages} {084052} (\bibinfo {year} {2017})},\ \Eprint
  {http://arxiv.org/abs/1710.09581} {arXiv:1710.09581 [gr-qc]} \BibitemShut
  {NoStop}%
\bibitem [{\citenamefont {Hu}\ and\ \citenamefont {Wu}(2017)}]{Hu:2017mde}%
  \BibitemOpen
  \bibfield  {author} {\bibinfo {author} {\bibfnamefont {W.-R.}\ \bibnamefont
  {Hu}}\ and\ \bibinfo {author} {\bibfnamefont {Y.-L.}\ \bibnamefont {Wu}},\
  }\href {\doibase 10.1093/nsr/nwx116} {\bibfield  {journal} {\bibinfo
  {journal} {Natl. Sci. Rev.}\ }\textbf {\bibinfo {volume} {4}},\ \bibinfo
  {pages} {685} (\bibinfo {year} {2017})}\BibitemShut {NoStop}%
\bibitem [{\citenamefont {Luo}\ \emph {et~al.}(2016)\citenamefont {Luo} \emph
  {et~al.}}]{TianQin:2015yph}%
  \BibitemOpen
  \bibfield  {author} {\bibinfo {author} {\bibfnamefont {J.}~\bibnamefont
  {Luo}} \emph {et~al.} (\bibinfo {collaboration} {TianQin}),\ }\href {\doibase
  10.1088/0264-9381/33/3/035010} {\bibfield  {journal} {\bibinfo  {journal}
  {Class. Quant. Grav.}\ }\textbf {\bibinfo {volume} {33}},\ \bibinfo {pages}
  {035010} (\bibinfo {year} {2016})},\ \Eprint
  {http://arxiv.org/abs/1512.02076} {arXiv:1512.02076 [astro-ph.IM]}
  \BibitemShut {NoStop}%
\bibitem [{\citenamefont {Gong}\ \emph {et~al.}(2021)\citenamefont {Gong},
  \citenamefont {Luo},\ and\ \citenamefont {Wang}}]{Gong:2021gvw}%
  \BibitemOpen
  \bibfield  {author} {\bibinfo {author} {\bibfnamefont {Y.}~\bibnamefont
  {Gong}}, \bibinfo {author} {\bibfnamefont {J.}~\bibnamefont {Luo}}, \ and\
  \bibinfo {author} {\bibfnamefont {B.}~\bibnamefont {Wang}},\ }\href {\doibase
  10.1038/s41550-021-01480-3} {\bibfield  {journal} {\bibinfo  {journal}
  {Nature Astron.}\ }\textbf {\bibinfo {volume} {5}},\ \bibinfo {pages} {881}
  (\bibinfo {year} {2021})},\ \Eprint {http://arxiv.org/abs/2109.07442}
  {arXiv:2109.07442 [astro-ph.IM]} \BibitemShut {NoStop}%
\bibitem [{\citenamefont {Danzmann}(1997)}]{Danzmann:1997hm}%
  \BibitemOpen
  \bibfield  {author} {\bibinfo {author} {\bibfnamefont {K.}~\bibnamefont
  {Danzmann}},\ }\href {\doibase 10.1088/0264-9381/14/6/002} {\bibfield
  {journal} {\bibinfo  {journal} {Class. Quant. Grav.}\ }\textbf {\bibinfo
  {volume} {14}},\ \bibinfo {pages} {1399} (\bibinfo {year}
  {1997})}\BibitemShut {NoStop}%
\bibitem [{\citenamefont {Schutz}(1999)}]{Schutz:1999xj}%
  \BibitemOpen
  \bibfield  {author} {\bibinfo {author} {\bibfnamefont {B.~F.}\ \bibnamefont
  {Schutz}},\ }\href {\doibase 10.1088/0264-9381/16/12A/307} {\bibfield
  {journal} {\bibinfo  {journal} {Class. Quant. Grav.}\ }\textbf {\bibinfo
  {volume} {16}},\ \bibinfo {pages} {A131} (\bibinfo {year} {1999})},\ \Eprint
  {http://arxiv.org/abs/gr-qc/9911034} {arXiv:gr-qc/9911034} \BibitemShut
  {NoStop}%
\bibitem [{\citenamefont {Gair}\ \emph {et~al.}(2004)\citenamefont {Gair},
  \citenamefont {Barack}, \citenamefont {Creighton}, \citenamefont {Cutler},
  \citenamefont {Larson}, \citenamefont {Phinney},\ and\ \citenamefont
  {Vallisneri}}]{Gair:2004iv}%
  \BibitemOpen
  \bibfield  {author} {\bibinfo {author} {\bibfnamefont {J.~R.}\ \bibnamefont
  {Gair}}, \bibinfo {author} {\bibfnamefont {L.}~\bibnamefont {Barack}},
  \bibinfo {author} {\bibfnamefont {T.}~\bibnamefont {Creighton}}, \bibinfo
  {author} {\bibfnamefont {C.}~\bibnamefont {Cutler}}, \bibinfo {author}
  {\bibfnamefont {S.~L.}\ \bibnamefont {Larson}}, \bibinfo {author}
  {\bibfnamefont {E.~S.}\ \bibnamefont {Phinney}}, \ and\ \bibinfo {author}
  {\bibfnamefont {M.}~\bibnamefont {Vallisneri}},\ }\href {\doibase
  10.1088/0264-9381/21/20/003} {\bibfield  {journal} {\bibinfo  {journal}
  {Class. Quant. Grav.}\ }\textbf {\bibinfo {volume} {21}},\ \bibinfo {pages}
  {S1595} (\bibinfo {year} {2004})},\ \Eprint
  {http://arxiv.org/abs/gr-qc/0405137} {arXiv:gr-qc/0405137} \BibitemShut
  {NoStop}%
\bibitem [{\citenamefont {Amaro-Seoane}\ \emph {et~al.}(2017)\citenamefont
  {Amaro-Seoane} \emph {et~al.}}]{LISA:2017pwj}%
  \BibitemOpen
  \bibfield  {author} {\bibinfo {author} {\bibfnamefont {P.}~\bibnamefont
  {Amaro-Seoane}} \emph {et~al.} (\bibinfo {collaboration} {LISA}),\
  }\href@noop {} {\  (\bibinfo {year} {2017})},\ \Eprint
  {http://arxiv.org/abs/1702.00786} {arXiv:1702.00786 [astro-ph.IM]}
  \BibitemShut {NoStop}%
\bibitem [{\citenamefont {Maselli}\ \emph {et~al.}(2022)\citenamefont
  {Maselli}, \citenamefont {Franchini}, \citenamefont {Gualtieri},
  \citenamefont {Sotiriou}, \citenamefont {Barsanti},\ and\ \citenamefont
  {Pani}}]{Maselli:2021men}%
  \BibitemOpen
  \bibfield  {author} {\bibinfo {author} {\bibfnamefont {A.}~\bibnamefont
  {Maselli}}, \bibinfo {author} {\bibfnamefont {N.}~\bibnamefont {Franchini}},
  \bibinfo {author} {\bibfnamefont {L.}~\bibnamefont {Gualtieri}}, \bibinfo
  {author} {\bibfnamefont {T.~P.}\ \bibnamefont {Sotiriou}}, \bibinfo {author}
  {\bibfnamefont {S.}~\bibnamefont {Barsanti}}, \ and\ \bibinfo {author}
  {\bibfnamefont {P.}~\bibnamefont {Pani}},\ }\href {\doibase
  10.1038/s41550-021-01589-5} {\bibfield  {journal} {\bibinfo  {journal}
  {Nature Astron.}\ }\textbf {\bibinfo {volume} {6}},\ \bibinfo {pages} {464}
  (\bibinfo {year} {2022})},\ \Eprint {http://arxiv.org/abs/2106.11325}
  {arXiv:2106.11325 [gr-qc]} \BibitemShut {NoStop}%
\bibitem [{\citenamefont {Bian}\ \emph {et~al.}(2026)\citenamefont {Bian} \emph
  {et~al.}}]{Bian:2025ifp}%
  \BibitemOpen
  \bibfield  {author} {\bibinfo {author} {\bibfnamefont {L.}~\bibnamefont
  {Bian}} \emph {et~al.},\ }\href {\doibase 10.1007/s11433-025-2740-8}
  {\bibfield  {journal} {\bibinfo  {journal} {Sci. China Phys. Mech. Astron.}\
  }\textbf {\bibinfo {volume} {69}},\ \bibinfo {pages} {210401} (\bibinfo
  {year} {2026})},\ \Eprint {http://arxiv.org/abs/2505.19747} {arXiv:2505.19747
  [gr-qc]} \BibitemShut {NoStop}%
\bibitem [{\citenamefont {Ni}(2024)}]{Ni:2024acg}%
  \BibitemOpen
  \bibfield  {author} {\bibinfo {author} {\bibfnamefont {W.-T.}\ \bibnamefont
  {Ni}},\ }\href {\doibase 10.1360/SSPMA-2024-0186} {\bibfield  {journal}
  {\bibinfo  {journal} {Sci. Sin. Phys. Mech. Astro.}\ }\textbf {\bibinfo
  {volume} {54}},\ \bibinfo {pages} {270402} (\bibinfo {year} {2024})},\
  \Eprint {http://arxiv.org/abs/2409.00927} {arXiv:2409.00927 [gr-qc]}
  \BibitemShut {NoStop}%
\bibitem [{\citenamefont {{Barausse}}\ \emph {et~al.}(2014)\citenamefont
  {{Barausse}}, \citenamefont {{Cardoso}},\ and\ \citenamefont
  {{Pani}}}]{Barausse14PRD}%
  \BibitemOpen
  \bibfield  {author} {\bibinfo {author} {\bibfnamefont {E.}~\bibnamefont
  {{Barausse}}}, \bibinfo {author} {\bibfnamefont {V.}~\bibnamefont
  {{Cardoso}}}, \ and\ \bibinfo {author} {\bibfnamefont {P.}~\bibnamefont
  {{Pani}}},\ }\href {\doibase 10.1103/PhysRevD.89.104059} {\bibfield
  {journal} {\bibinfo  {journal} {Phys. Rev. D}\ }\textbf {\bibinfo {volume}
  {89}},\ \bibinfo {eid} {104059} (\bibinfo {year} {2014})},\ \Eprint
  {http://arxiv.org/abs/1404.7149} {arXiv:1404.7149 [gr-qc]} \BibitemShut
  {NoStop}%
\bibitem [{\citenamefont {{Cardoso}}\ \emph {et~al.}(2022)\citenamefont
  {{Cardoso}}, \citenamefont {{Destounis}}, \citenamefont {{Duque}},
  \citenamefont {{Macedo}},\ and\ \citenamefont {{Maselli}}}]{Cardoso22PRD}%
  \BibitemOpen
  \bibfield  {author} {\bibinfo {author} {\bibfnamefont {V.}~\bibnamefont
  {{Cardoso}}}, \bibinfo {author} {\bibfnamefont {K.}~\bibnamefont
  {{Destounis}}}, \bibinfo {author} {\bibfnamefont {F.}~\bibnamefont
  {{Duque}}}, \bibinfo {author} {\bibfnamefont {R.~P.}\ \bibnamefont
  {{Macedo}}}, \ and\ \bibinfo {author} {\bibfnamefont {A.}~\bibnamefont
  {{Maselli}}},\ }\href {\doibase 10.1103/PhysRevD.105.L061501} {\bibfield
  {journal} {\bibinfo  {journal} {Phys. Rev. D}\ }\textbf {\bibinfo {volume}
  {105}},\ \bibinfo {eid} {L061501} (\bibinfo {year} {2022})},\ \Eprint
  {http://arxiv.org/abs/2109.00005} {arXiv:2109.00005 [gr-qc]} \BibitemShut
  {NoStop}%
\bibitem [{\citenamefont {Zhang}\ and\ \citenamefont
  {Zhu}(2026)}]{Zhang26EPJC}%
  \BibitemOpen
  \bibfield  {author} {\bibinfo {author} {\bibfnamefont {C.}~\bibnamefont
  {Zhang}}\ and\ \bibinfo {author} {\bibfnamefont {T.}~\bibnamefont {Zhu}},\
  }\href {\doibase 10.1140/epjc/s10052-026-15606-2} {\bibfield  {journal}
  {\bibinfo  {journal} {Eur. Phys. J. C}\ }\textbf {\bibinfo {volume} {86}}
  (\bibinfo {year} {2026}),\ 10.1140/epjc/s10052-026-15606-2}\BibitemShut
  {NoStop}%
\bibitem [{\citenamefont {Yang}\ \emph {et~al.}(2025)\citenamefont {Yang},
  \citenamefont {Zhang}, \citenamefont {Zhu}, \citenamefont {Zhao},\ and\
  \citenamefont {Liu}}]{Yang24JCAP}%
  \BibitemOpen
  \bibfield  {author} {\bibinfo {author} {\bibfnamefont {S.}~\bibnamefont
  {Yang}}, \bibinfo {author} {\bibfnamefont {Y.-P.}\ \bibnamefont {Zhang}},
  \bibinfo {author} {\bibfnamefont {T.}~\bibnamefont {Zhu}}, \bibinfo {author}
  {\bibfnamefont {L.}~\bibnamefont {Zhao}}, \ and\ \bibinfo {author}
  {\bibfnamefont {Y.-X.}\ \bibnamefont {Liu}},\ }\href {\doibase
  10.1088/1475-7516/2025/01/091} {\bibfield  {journal} {\bibinfo  {journal}
  {JCAP}\ }\textbf {\bibinfo {volume} {01}},\ \bibinfo {pages} {091} (\bibinfo
  {year} {2025})},\ \Eprint {http://arxiv.org/abs/2407.00283} {arXiv:2407.00283
  [gr-qc]} \BibitemShut {NoStop}%
\bibitem [{\citenamefont {Shabbir}\ \emph {et~al.}(2025)\citenamefont
  {Shabbir}, \citenamefont {Jamil},\ and\ \citenamefont
  {Azreg-A{\"\i}nou}}]{Shabbir25}%
  \BibitemOpen
  \bibfield  {author} {\bibinfo {author} {\bibfnamefont {O.}~\bibnamefont
  {Shabbir}}, \bibinfo {author} {\bibfnamefont {M.}~\bibnamefont {Jamil}}, \
  and\ \bibinfo {author} {\bibfnamefont {M.}~\bibnamefont {Azreg-A{\"\i}nou}},\
  }\href {\doibase 10.1016/j.dark.2025.101816} {\bibfield  {journal} {\bibinfo
  {journal} {Phys. Dark Univ.}\ }\textbf {\bibinfo {volume} {47}},\ \bibinfo
  {pages} {101816} (\bibinfo {year} {2025})},\ \Eprint
  {http://arxiv.org/abs/2501.04367} {arXiv:2501.04367 [gr-qc]} \BibitemShut
  {NoStop}%
\bibitem [{\citenamefont {Junior}\ \emph {et~al.}(2025)\citenamefont {Junior},
  \citenamefont {Junior}, \citenamefont {Lobo}, \citenamefont {Rodrigues},
  \citenamefont {Rubiera-Garcia}, \citenamefont {da~Silva},\ and\ \citenamefont
  {Vieira}}]{Junior24}%
  \BibitemOpen
  \bibfield  {author} {\bibinfo {author} {\bibfnamefont {E.~L.~B.}\
  \bibnamefont {Junior}}, \bibinfo {author} {\bibfnamefont {J.~T. S.~S.}\
  \bibnamefont {Junior}}, \bibinfo {author} {\bibfnamefont {F.~S.~N.}\
  \bibnamefont {Lobo}}, \bibinfo {author} {\bibfnamefont {M.~E.}\ \bibnamefont
  {Rodrigues}}, \bibinfo {author} {\bibfnamefont {D.}~\bibnamefont
  {Rubiera-Garcia}}, \bibinfo {author} {\bibfnamefont {L.~F.~D.}\ \bibnamefont
  {da~Silva}}, \ and\ \bibinfo {author} {\bibfnamefont {H.~A.}\ \bibnamefont
  {Vieira}},\ }\href {\doibase 10.1140/epjc/s10052-025-14299-3} {\bibfield
  {journal} {\bibinfo  {journal} {Eur. Phys. J. C}\ }\textbf {\bibinfo {volume}
  {85}},\ \bibinfo {pages} {557} (\bibinfo {year} {2025})},\ \Eprint
  {http://arxiv.org/abs/2412.00769} {arXiv:2412.00769 [gr-qc]} \BibitemShut
  {NoStop}%
\bibitem [{\citenamefont {Haroon}\ and\ \citenamefont {Zhu}(2025)}]{Haroon25}%
  \BibitemOpen
  \bibfield  {author} {\bibinfo {author} {\bibfnamefont {S.}~\bibnamefont
  {Haroon}}\ and\ \bibinfo {author} {\bibfnamefont {T.}~\bibnamefont {Zhu}},\
  }\href {\doibase 10.1103/ckdt-wtsl} {\bibfield  {journal} {\bibinfo
  {journal} {Phys. Rev. D}\ }\textbf {\bibinfo {volume} {112}},\ \bibinfo
  {pages} {044046} (\bibinfo {year} {2025})},\ \Eprint
  {http://arxiv.org/abs/2502.09171} {arXiv:2502.09171 [gr-qc]} \BibitemShut
  {NoStop}%
\bibitem [{\citenamefont {Alloqulov}\ \emph {et~al.}(2025)\citenamefont
  {Alloqulov}, \citenamefont {Xamidov}, \citenamefont {Shaymatov},\ and\
  \citenamefont {Ahmedov}}]{Alloqulov25GW}%
  \BibitemOpen
  \bibfield  {author} {\bibinfo {author} {\bibfnamefont {M.}~\bibnamefont
  {Alloqulov}}, \bibinfo {author} {\bibfnamefont {T.}~\bibnamefont {Xamidov}},
  \bibinfo {author} {\bibfnamefont {S.}~\bibnamefont {Shaymatov}}, \ and\
  \bibinfo {author} {\bibfnamefont {B.}~\bibnamefont {Ahmedov}},\ }\href
  {\doibase 10.1140/epjc/s10052-025-14529-8} {\bibfield  {journal} {\bibinfo
  {journal} {Eur. Phys. J. C}\ }\textbf {\bibinfo {volume} {85}},\ \bibinfo
  {pages} {798} (\bibinfo {year} {2025})},\ \Eprint
  {http://arxiv.org/abs/2504.05236} {arXiv:2504.05236 [gr-qc]} \BibitemShut
  {NoStop}%
\bibitem [{\citenamefont {Wang}\ \emph {et~al.}(2025)\citenamefont {Wang},
  \citenamefont {Meng}, \citenamefont {Zhang}, \citenamefont {Zhu},\ and\
  \citenamefont {Wei}}]{Wang25JCAP}%
  \BibitemOpen
  \bibfield  {author} {\bibinfo {author} {\bibfnamefont {C.-H.}\ \bibnamefont
  {Wang}}, \bibinfo {author} {\bibfnamefont {X.-C.}\ \bibnamefont {Meng}},
  \bibinfo {author} {\bibfnamefont {Y.-P.}\ \bibnamefont {Zhang}}, \bibinfo
  {author} {\bibfnamefont {T.}~\bibnamefont {Zhu}}, \ and\ \bibinfo {author}
  {\bibfnamefont {S.-W.}\ \bibnamefont {Wei}},\ }\href {\doibase
  10.1088/1475-7516/2025/07/021} {\bibfield  {journal} {\bibinfo  {journal} {J.
  Cosmol. Astropart. Phys.}\ }\textbf {\bibinfo {volume} {2025}},\ \bibinfo
  {pages} {021} (\bibinfo {year} {2025})}\BibitemShut {NoStop}%
\bibitem [{\citenamefont {Lu}\ and\ \citenamefont {Zhu}(2025)}]{Lu25GW}%
  \BibitemOpen
  \bibfield  {author} {\bibinfo {author} {\bibfnamefont {S.}~\bibnamefont
  {Lu}}\ and\ \bibinfo {author} {\bibfnamefont {T.}~\bibnamefont {Zhu}},\
  }\href {\doibase 10.1016/j.dark.2025.102141} {\bibfield  {journal} {\bibinfo
  {journal} {Phys. Dark Universe}\ }\textbf {\bibinfo {volume} {50}},\ \bibinfo
  {pages} {102141} (\bibinfo {year} {2025})},\ \Eprint
  {http://arxiv.org/abs/2505.00294} {arXiv:2505.00294 [gr-qc]} \BibitemShut
  {NoStop}%
\bibitem [{\citenamefont {Ahmed}\ \emph {et~al.}(2026)\citenamefont {Ahmed},
  \citenamefont {Wu}, \citenamefont {Ghosh},\ and\ \citenamefont
  {Zhu}}]{Ahmed26GW1}%
  \BibitemOpen
  \bibfield  {author} {\bibinfo {author} {\bibfnamefont {F.}~\bibnamefont
  {Ahmed}}, \bibinfo {author} {\bibfnamefont {Q.}~\bibnamefont {Wu}}, \bibinfo
  {author} {\bibfnamefont {S.~G.}\ \bibnamefont {Ghosh}}, \ and\ \bibinfo
  {author} {\bibfnamefont {T.}~\bibnamefont {Zhu}},\ }\href {\doibase
  10.1088/1475-7516/2026/02/004} {\bibfield  {journal} {\bibinfo  {journal} {J.
  Cosmol. Astropart. Phys.}\ }\textbf {\bibinfo {volume} {2026}},\ \bibinfo
  {pages} {004} (\bibinfo {year} {2026})}\BibitemShut {NoStop}%
\bibitem [{\citenamefont {Ahmed}\ \emph {et~al.}(2025)\citenamefont {Ahmed},
  \citenamefont {Wu}, \citenamefont {Ghosh},\ and\ \citenamefont
  {Zhu}}]{Ahmed25GW2}%
  \BibitemOpen
  \bibfield  {author} {\bibinfo {author} {\bibfnamefont {F.}~\bibnamefont
  {Ahmed}}, \bibinfo {author} {\bibfnamefont {Q.}~\bibnamefont {Wu}}, \bibinfo
  {author} {\bibfnamefont {S.~G.}\ \bibnamefont {Ghosh}}, \ and\ \bibinfo
  {author} {\bibfnamefont {T.}~\bibnamefont {Zhu}},\ }\href
  {https://arxiv.org/abs/2512.24036} {\enquote {\bibinfo {title} {Signatures of
  quantum-corrected black holes in gravitational waves from periodic orbits},}\
  } (\bibinfo {year} {2025}),\ \Eprint {http://arxiv.org/abs/2512.24036}
  {arXiv:2512.24036 [gr-qc]} \BibitemShut {NoStop}%
\bibitem [{\citenamefont {Healy}\ \emph {et~al.}(2009)\citenamefont {Healy},
  \citenamefont {Levin},\ and\ \citenamefont {Shoemaker}}]{Healy09PRL}%
  \BibitemOpen
  \bibfield  {author} {\bibinfo {author} {\bibfnamefont {J.}~\bibnamefont
  {Healy}}, \bibinfo {author} {\bibfnamefont {J.}~\bibnamefont {Levin}}, \ and\
  \bibinfo {author} {\bibfnamefont {D.}~\bibnamefont {Shoemaker}},\ }\href
  {\doibase 10.1103/PhysRevLett.103.131101} {\bibfield  {journal} {\bibinfo
  {journal} {Phys. Rev. Lett.}\ }\textbf {\bibinfo {volume} {103}},\ \bibinfo
  {pages} {131101} (\bibinfo {year} {2009})},\ \Eprint
  {http://arxiv.org/abs/0907.0671} {arXiv:0907.0671 [gr-qc]} \BibitemShut
  {NoStop}%
\bibitem [{\citenamefont {Pugliese}\ \emph {et~al.}(2017)\citenamefont
  {Pugliese}, \citenamefont {Quevedo},\ and\ \citenamefont
  {Ruffini}}]{Pugliese13}%
  \BibitemOpen
  \bibfield  {author} {\bibinfo {author} {\bibfnamefont {D.}~\bibnamefont
  {Pugliese}}, \bibinfo {author} {\bibfnamefont {H.}~\bibnamefont {Quevedo}}, \
  and\ \bibinfo {author} {\bibfnamefont {R.}~\bibnamefont {Ruffini}},\ }\href
  {\doibase 10.1140/epjc/s10052-017-4769-x} {\bibfield  {journal} {\bibinfo
  {journal} {Eur. Phys. J. C}\ }\textbf {\bibinfo {volume} {77}},\ \bibinfo
  {pages} {206} (\bibinfo {year} {2017})},\ \Eprint
  {http://arxiv.org/abs/1304.2940} {arXiv:1304.2940 [gr-qc]} \BibitemShut
  {NoStop}%
\bibitem [{\citenamefont {Lin}\ and\ \citenamefont {Deng}(2023)}]{Lin23}%
  \BibitemOpen
  \bibfield  {author} {\bibinfo {author} {\bibfnamefont {H.-Y.}\ \bibnamefont
  {Lin}}\ and\ \bibinfo {author} {\bibfnamefont {X.-M.}\ \bibnamefont {Deng}},\
  }\href {\doibase 10.1140/epjc/s10052-023-11487-x} {\bibfield  {journal}
  {\bibinfo  {journal} {Eur. Phys. J. C}\ }\textbf {\bibinfo {volume} {83}},\
  \bibinfo {pages} {311} (\bibinfo {year} {2023})}\BibitemShut {NoStop}%
\bibitem [{\citenamefont {Yao}\ and\ \citenamefont {Li}(2023)}]{Yao23}%
  \BibitemOpen
  \bibfield  {author} {\bibinfo {author} {\bibfnamefont {J.-T.}\ \bibnamefont
  {Yao}}\ and\ \bibinfo {author} {\bibfnamefont {X.}~\bibnamefont {Li}},\
  }\href {\doibase 10.1103/PhysRevD.108.084067} {\bibfield  {journal} {\bibinfo
   {journal} {Phys. Rev. D}\ }\textbf {\bibinfo {volume} {108}},\ \bibinfo
  {pages} {084067} (\bibinfo {year} {2023})}\BibitemShut {NoStop}%
\bibitem [{\citenamefont {Lin}\ and\ \citenamefont {Deng}(2021)}]{Lin21}%
  \BibitemOpen
  \bibfield  {author} {\bibinfo {author} {\bibfnamefont {H.-Y.}\ \bibnamefont
  {Lin}}\ and\ \bibinfo {author} {\bibfnamefont {X.-M.}\ \bibnamefont {Deng}},\
  }\href {\doibase 10.1016/j.dark.2020.100745} {\bibfield  {journal} {\bibinfo
  {journal} {Phys. Dark Univ.}\ }\textbf {\bibinfo {volume} {31}},\ \bibinfo
  {pages} {100745} (\bibinfo {year} {2021})}\BibitemShut {NoStop}%
\bibitem [{\citenamefont {Tu}\ \emph {et~al.}(2023)\citenamefont {Tu},
  \citenamefont {Zhu},\ and\ \citenamefont {Wang}}]{Tu23}%
  \BibitemOpen
  \bibfield  {author} {\bibinfo {author} {\bibfnamefont {Z.-Y.}\ \bibnamefont
  {Tu}}, \bibinfo {author} {\bibfnamefont {T.}~\bibnamefont {Zhu}}, \ and\
  \bibinfo {author} {\bibfnamefont {A.}~\bibnamefont {Wang}},\ }\href {\doibase
  10.1103/PhysRevD.108.024035} {\bibfield  {journal} {\bibinfo  {journal}
  {Phys. Rev. D}\ }\textbf {\bibinfo {volume} {108}},\ \bibinfo {pages}
  {024035} (\bibinfo {year} {2023})}\BibitemShut {NoStop}%
\bibitem [{\citenamefont {Deng}(2020)}]{Deng20}%
  \BibitemOpen
  \bibfield  {author} {\bibinfo {author} {\bibfnamefont {X.-M.}\ \bibnamefont
  {Deng}},\ }\href {\doibase 10.1140/epjc/s10052-020-8067-7} {\bibfield
  {journal} {\bibinfo  {journal} {Eur. Phys. J. C}\ }\textbf {\bibinfo {volume}
  {80}},\ \bibinfo {pages} {489} (\bibinfo {year} {2020})}\BibitemShut
  {NoStop}%
\bibitem [{\citenamefont {Wei}\ \emph {et~al.}(2019)\citenamefont {Wei},
  \citenamefont {Yang},\ and\ \citenamefont {Liu}}]{Wei19}%
  \BibitemOpen
  \bibfield  {author} {\bibinfo {author} {\bibfnamefont {S.-W.}\ \bibnamefont
  {Wei}}, \bibinfo {author} {\bibfnamefont {J.}~\bibnamefont {Yang}}, \ and\
  \bibinfo {author} {\bibfnamefont {Y.-X.}\ \bibnamefont {Liu}},\ }\href
  {\doibase 10.1103/PhysRevD.99.104016} {\bibfield  {journal} {\bibinfo
  {journal} {Phys. Rev. D}\ }\textbf {\bibinfo {volume} {99}},\ \bibinfo
  {pages} {104016} (\bibinfo {year} {2019})},\ \Eprint
  {http://arxiv.org/abs/1904.03129} {arXiv:1904.03129 [gr-qc]} \BibitemShut
  {NoStop}%
\bibitem [{\citenamefont {Jiang}\ \emph {et~al.}(2024)\citenamefont {Jiang},
  \citenamefont {Alloqulov}, \citenamefont {Wu}, \citenamefont {Shaymatov},\
  and\ \citenamefont {Zhu}}]{JIANG2024}%
  \BibitemOpen
  \bibfield  {author} {\bibinfo {author} {\bibfnamefont {H.}~\bibnamefont
  {Jiang}}, \bibinfo {author} {\bibfnamefont {M.}~\bibnamefont {Alloqulov}},
  \bibinfo {author} {\bibfnamefont {Q.}~\bibnamefont {Wu}}, \bibinfo {author}
  {\bibfnamefont {S.}~\bibnamefont {Shaymatov}}, \ and\ \bibinfo {author}
  {\bibfnamefont {T.}~\bibnamefont {Zhu}},\ }\href {\doibase
  https://doi.org/10.1016/j.dark.2024.101627} {\bibfield  {journal} {\bibinfo
  {journal} {Phys. Dark Universe}\ }\textbf {\bibinfo {volume} {46}},\ \bibinfo
  {pages} {101627} (\bibinfo {year} {2024})}\BibitemShut {NoStop}%
\bibitem [{\citenamefont {Wei}\ \emph {et~al.}(2025)\citenamefont {Wei},
  \citenamefont {Zhang}, \citenamefont {Xie},\ and\ \citenamefont
  {Yin}}]{Wei25}%
  \BibitemOpen
  \bibfield  {author} {\bibinfo {author} {\bibfnamefont {Z.-L.}\ \bibnamefont
  {Wei}}, \bibinfo {author} {\bibfnamefont {J.}~\bibnamefont {Zhang}}, \bibinfo
  {author} {\bibfnamefont {Y.}~\bibnamefont {Xie}}, \ and\ \bibinfo {author}
  {\bibfnamefont {P.-L.}\ \bibnamefont {Yin}},\ }\href {\doibase
  10.1140/epjc/s10052-025-14437-x} {\bibfield  {journal} {\bibinfo  {journal}
  {Eur. Phys. J. C}\ }\textbf {\bibinfo {volume} {85}},\ \bibinfo {pages} {698}
  (\bibinfo {year} {2025})}\BibitemShut {NoStop}%
\bibitem [{\citenamefont {Alloqulov}\ \emph {et~al.}(2026)\citenamefont
  {Alloqulov}, \citenamefont {Shaymatov}, \citenamefont {Ahmedov},\ and\
  \citenamefont {Zhu}}]{Alloqulov26GW1}%
  \BibitemOpen
  \bibfield  {author} {\bibinfo {author} {\bibfnamefont {M.}~\bibnamefont
  {Alloqulov}}, \bibinfo {author} {\bibfnamefont {S.}~\bibnamefont
  {Shaymatov}}, \bibinfo {author} {\bibfnamefont {B.}~\bibnamefont {Ahmedov}},
  \ and\ \bibinfo {author} {\bibfnamefont {T.}~\bibnamefont {Zhu}},\ }\href
  {\doibase 10.1140/epjc/s10052-025-15251-1} {\bibfield  {journal} {\bibinfo
  {journal} {Eur. Phys. J. C}\ }\textbf {\bibinfo {volume} {86}},\ \bibinfo
  {pages} {117} (\bibinfo {year} {2026})}\BibitemShut {NoStop}%
\bibitem [{\citenamefont {Sharipov}\ \emph {et~al.}(2025)\citenamefont
  {Sharipov}, \citenamefont {Xamidov}, \citenamefont {Wu}, \citenamefont
  {Shaymatov},\ and\ \citenamefont {Zhu}}]{Sharipov25}%
  \BibitemOpen
  \bibfield  {author} {\bibinfo {author} {\bibfnamefont {J.}~\bibnamefont
  {Sharipov}}, \bibinfo {author} {\bibfnamefont {T.}~\bibnamefont {Xamidov}},
  \bibinfo {author} {\bibfnamefont {Q.}~\bibnamefont {Wu}}, \bibinfo {author}
  {\bibfnamefont {S.}~\bibnamefont {Shaymatov}}, \ and\ \bibinfo {author}
  {\bibfnamefont {T.}~\bibnamefont {Zhu}},\ }\href@noop {} {\  (\bibinfo {year}
  {2025})},\ \Eprint {http://arxiv.org/abs/2511.10043} {arXiv:2511.10043
  [gr-qc]} \BibitemShut {NoStop}%
\bibitem [{\citenamefont {Ginzburg V.~L.}(1964)}]{Ginzburg1964}%
  \BibitemOpen
  \bibfield  {author} {\bibinfo {author} {\bibfnamefont {O.~L.~M.}\
  \bibnamefont {Ginzburg V.~L.}},\ }\href@noop {} {\bibfield  {journal}
  {\bibinfo  {journal} {Zh. Eksp. Teor. Fiz.}\ }\textbf {\bibinfo {volume}
  {47}},\ \bibinfo {pages} {1030} (\bibinfo {year} {1964})}\BibitemShut
  {NoStop}%
\bibitem [{\citenamefont {{Anderson}}\ and\ \citenamefont
  {{Cohen}}(1970)}]{Anderson70}%
  \BibitemOpen
  \bibfield  {author} {\bibinfo {author} {\bibfnamefont {J.~L.}\ \bibnamefont
  {{Anderson}}}\ and\ \bibinfo {author} {\bibfnamefont {J.~M.}\ \bibnamefont
  {{Cohen}}},\ }\href {\doibase 10.1007/BF00649960} {\bibfield  {journal}
  {\bibinfo  {journal} {Astrophys. Space Sci.}\ }\textbf {\bibinfo {volume}
  {9}},\ \bibinfo {pages} {146} (\bibinfo {year} {1970})}\BibitemShut {NoStop}%
\bibitem [{\citenamefont {{Wald}}(1974)}]{Wald74}%
  \BibitemOpen
  \bibfield  {author} {\bibinfo {author} {\bibfnamefont {R.~M.}\ \bibnamefont
  {{Wald}}},\ }\href {\doibase 10.1103/PhysRevD.10.1680} {\bibfield  {journal}
  {\bibinfo  {journal} {Phys. Rev. D}\ }\textbf {\bibinfo {volume} {10}},\
  \bibinfo {pages} {1680} (\bibinfo {year} {1974})}\BibitemShut {NoStop}%
\bibitem [{\citenamefont {{Rezzolla}}\ \emph {et~al.}(2001)\citenamefont
  {{Rezzolla}}, \citenamefont {{Ahmedov}},\ and\ \citenamefont
  {{Miller}}}]{Rezzolla01}%
  \BibitemOpen
  \bibfield  {author} {\bibinfo {author} {\bibfnamefont {L.}~\bibnamefont
  {{Rezzolla}}}, \bibinfo {author} {\bibfnamefont {B.~J.}\ \bibnamefont
  {{Ahmedov}}}, \ and\ \bibinfo {author} {\bibfnamefont {J.~C.}\ \bibnamefont
  {{Miller}}},\ }\href {\doibase 10.1046/j.1365-8711.2001.04161.x} {\bibfield
  {journal} {\bibinfo  {journal} {Mon. Not. R. Astron. Soc.}\ }\textbf
  {\bibinfo {volume} {322}},\ \bibinfo {pages} {723} (\bibinfo {year}
  {2001})},\ \Eprint {http://arxiv.org/abs/astro-ph/0011316}
  {arXiv:astro-ph/0011316 [astro-ph]} \BibitemShut {NoStop}%
\bibitem [{\citenamefont {{de Felice}}\ and\ \citenamefont
  {{Sorge}}(2003)}]{deFelice03}%
  \BibitemOpen
  \bibfield  {author} {\bibinfo {author} {\bibfnamefont {F.}~\bibnamefont {{de
  Felice}}}\ and\ \bibinfo {author} {\bibfnamefont {F.}~\bibnamefont
  {{Sorge}}},\ }\href@noop {} {\bibfield  {journal} {\bibinfo  {journal}
  {Class. Quantum Grav.}\ }\textbf {\bibinfo {volume} {20}},\ \bibinfo {pages}
  {469} (\bibinfo {year} {2003})}\BibitemShut {NoStop}%
\bibitem [{\citenamefont {{Frolov}}\ and\ \citenamefont
  {{Shoom}}(2010)}]{Frolov10}%
  \BibitemOpen
  \bibfield  {author} {\bibinfo {author} {\bibfnamefont {V.~P.}\ \bibnamefont
  {{Frolov}}}\ and\ \bibinfo {author} {\bibfnamefont {A.~A.}\ \bibnamefont
  {{Shoom}}},\ }\href {\doibase 10.1103/PhysRevD.82.084034} {\bibfield
  {journal} {\bibinfo  {journal} {Phys. Rev. D}\ }\textbf {\bibinfo {volume}
  {82}},\ \bibinfo {eid} {084034} (\bibinfo {year} {2010})},\ \Eprint
  {http://arxiv.org/abs/1008.2985} {arXiv:1008.2985 [gr-qc]} \BibitemShut
  {NoStop}%
\bibitem [{\citenamefont {{Aliev}}\ and\ \citenamefont
  {{{\"O}zdemir}}(2002)}]{Aliev02}%
  \BibitemOpen
  \bibfield  {author} {\bibinfo {author} {\bibfnamefont {A.~N.}\ \bibnamefont
  {{Aliev}}}\ and\ \bibinfo {author} {\bibfnamefont {N.}~\bibnamefont
  {{{\"O}zdemir}}},\ }\href {\doibase 10.1046/j.1365-8711.2002.05727.x}
  {\bibfield  {journal} {\bibinfo  {journal} {Mon. Not. R. Astron. Soc.}\
  }\textbf {\bibinfo {volume} {336}},\ \bibinfo {pages} {241} (\bibinfo {year}
  {2002})},\ \Eprint {http://arxiv.org/abs/gr-qc/0208025} {gr-qc/0208025}
  \BibitemShut {NoStop}%
\bibitem [{\citenamefont {{Abdujabbarov}}\ and\ \citenamefont
  {{Ahmedov}}(2010)}]{Abdujabbarov10}%
  \BibitemOpen
  \bibfield  {author} {\bibinfo {author} {\bibfnamefont {A.}~\bibnamefont
  {{Abdujabbarov}}}\ and\ \bibinfo {author} {\bibfnamefont {B.}~\bibnamefont
  {{Ahmedov}}},\ }\href {\doibase 10.1103/PhysRevD.81.044022} {\bibfield
  {journal} {\bibinfo  {journal} {Phys. Rev. D}\ }\textbf {\bibinfo {volume}
  {81}},\ \bibinfo {eid} {044022} (\bibinfo {year} {2010})},\ \Eprint
  {http://arxiv.org/abs/0905.2730} {arXiv:0905.2730 [gr-qc]} \BibitemShut
  {NoStop}%
\bibitem [{\citenamefont {{Shaymatov}}\ \emph {et~al.}(2014)\citenamefont
  {{Shaymatov}}, \citenamefont {{Atamurotov}},\ and\ \citenamefont
  {{Ahmedov}}}]{Shaymatov14}%
  \BibitemOpen
  \bibfield  {author} {\bibinfo {author} {\bibfnamefont {S.}~\bibnamefont
  {{Shaymatov}}}, \bibinfo {author} {\bibfnamefont {F.}~\bibnamefont
  {{Atamurotov}}}, \ and\ \bibinfo {author} {\bibfnamefont {B.}~\bibnamefont
  {{Ahmedov}}},\ }\href {\doibase 10.1007/s10509-013-1752-3} {\bibfield
  {journal} {\bibinfo  {journal} {Astrophys Space Sci}\ }\textbf {\bibinfo
  {volume} {350}},\ \bibinfo {pages} {413} (\bibinfo {year}
  {2014})}\BibitemShut {NoStop}%
\bibitem [{\citenamefont {{Jamil}}\ \emph {et~al.}(2015)\citenamefont
  {{Jamil}}, \citenamefont {{Hussain}},\ and\ \citenamefont
  {{Majeed}}}]{Jamil15}%
  \BibitemOpen
  \bibfield  {author} {\bibinfo {author} {\bibfnamefont {M.}~\bibnamefont
  {{Jamil}}}, \bibinfo {author} {\bibfnamefont {S.}~\bibnamefont {{Hussain}}},
  \ and\ \bibinfo {author} {\bibfnamefont {B.}~\bibnamefont {{Majeed}}},\
  }\href {\doibase 10.1140/epjc/s10052-014-3230-7} {\bibfield  {journal}
  {\bibinfo  {journal} {Eur. Phys. J. C}\ }\textbf {\bibinfo {volume} {75}},\
  \bibinfo {eid} {24} (\bibinfo {year} {2015})},\ \Eprint
  {http://arxiv.org/abs/1404.7123} {arXiv:1404.7123 [gr-qc]} \BibitemShut
  {NoStop}%
\bibitem [{\citenamefont {{Tursunov}}\ \emph {et~al.}(2016)\citenamefont
  {{Tursunov}}, \citenamefont {{Stuchl{\'{\i}}k}},\ and\ \citenamefont
  {{Kolo{\v s}}}}]{Tursunov16}%
  \BibitemOpen
  \bibfield  {author} {\bibinfo {author} {\bibfnamefont {A.}~\bibnamefont
  {{Tursunov}}}, \bibinfo {author} {\bibfnamefont {Z.}~\bibnamefont
  {{Stuchl{\'{\i}}k}}}, \ and\ \bibinfo {author} {\bibfnamefont
  {M.}~\bibnamefont {{Kolo{\v s}}}},\ }\href {\doibase
  10.1103/PhysRevD.93.084012} {\bibfield  {journal} {\bibinfo  {journal} {Phys.
  Rev. D}\ }\textbf {\bibinfo {volume} {93}},\ \bibinfo {eid} {084012}
  (\bibinfo {year} {2016})},\ \Eprint {http://arxiv.org/abs/1603.07264}
  {arXiv:1603.07264 [gr-qc]} \BibitemShut {NoStop}%
\bibitem [{\citenamefont {{Hussain}}\ and\ \citenamefont
  {{Jamil}}(2015)}]{Hussain15}%
  \BibitemOpen
  \bibfield  {author} {\bibinfo {author} {\bibfnamefont {S.}~\bibnamefont
  {{Hussain}}}\ and\ \bibinfo {author} {\bibfnamefont {M.}~\bibnamefont
  {{Jamil}}},\ }\href {\doibase 10.1103/PhysRevD.92.043008} {\bibfield
  {journal} {\bibinfo  {journal} {Phys. Rev. D}\ }\textbf {\bibinfo {volume}
  {92}},\ \bibinfo {eid} {043008} (\bibinfo {year} {2015})},\ \Eprint
  {http://arxiv.org/abs/1508.02123} {arXiv:1508.02123 [gr-qc]} \BibitemShut
  {NoStop}%
\bibitem [{\citenamefont {{Shaymatov}}\ \emph {et~al.}(2015)\citenamefont
  {{Shaymatov}}, \citenamefont {{Patil}}, \citenamefont {{Ahmedov}},\ and\
  \citenamefont {{Joshi}}}]{Shaymatov15}%
  \BibitemOpen
  \bibfield  {author} {\bibinfo {author} {\bibfnamefont {S.}~\bibnamefont
  {{Shaymatov}}}, \bibinfo {author} {\bibfnamefont {M.}~\bibnamefont
  {{Patil}}}, \bibinfo {author} {\bibfnamefont {B.}~\bibnamefont {{Ahmedov}}},
  \ and\ \bibinfo {author} {\bibfnamefont {P.~S.}\ \bibnamefont {{Joshi}}},\
  }\href {\doibase 10.1103/PhysRevD.91.064025} {\bibfield  {journal} {\bibinfo
  {journal} {Phys. Rev. D}\ }\textbf {\bibinfo {volume} {91}},\ \bibinfo {eid}
  {064025} (\bibinfo {year} {2015})},\ \Eprint {http://arxiv.org/abs/1409.3018}
  {arXiv:1409.3018 [gr-qc]} \BibitemShut {NoStop}%
\bibitem [{\citenamefont {{Shaymatov}}\ \emph
  {et~al.}(2021{\natexlab{a}})\citenamefont {{Shaymatov}}, \citenamefont
  {{Malafarina}},\ and\ \citenamefont {{Ahmedov}}}]{Shaymatov21pdu}%
  \BibitemOpen
  \bibfield  {author} {\bibinfo {author} {\bibfnamefont {S.}~\bibnamefont
  {{Shaymatov}}}, \bibinfo {author} {\bibfnamefont {D.}~\bibnamefont
  {{Malafarina}}}, \ and\ \bibinfo {author} {\bibfnamefont {B.}~\bibnamefont
  {{Ahmedov}}},\ }\href {\doibase 10.1016/j.dark.2021.100891} {\bibfield
  {journal} {\bibinfo  {journal} {Phys. Dark Universe}\ }\textbf {\bibinfo
  {volume} {34}},\ \bibinfo {eid} {100891} (\bibinfo {year}
  {2021}{\natexlab{a}})},\ \Eprint {http://arxiv.org/abs/2004.06811}
  {arXiv:2004.06811 [gr-qc]} \BibitemShut {NoStop}%
\bibitem [{\citenamefont {{Shaymatov}}\ \emph
  {et~al.}(2021{\natexlab{b}})\citenamefont {{Shaymatov}}, \citenamefont
  {{Ahmedov}},\ and\ \citenamefont {{Jamil}}}]{Shaymatov21d}%
  \BibitemOpen
  \bibfield  {author} {\bibinfo {author} {\bibfnamefont {S.}~\bibnamefont
  {{Shaymatov}}}, \bibinfo {author} {\bibfnamefont {B.}~\bibnamefont
  {{Ahmedov}}}, \ and\ \bibinfo {author} {\bibfnamefont {M.}~\bibnamefont
  {{Jamil}}},\ }\href {\doibase 10.1140/epjc/s10052-021-09398-w} {\bibfield
  {journal} {\bibinfo  {journal} {Eur. Phys. J. C}\ }\textbf {\bibinfo {volume}
  {81}},\ \bibinfo {eid} {588} (\bibinfo {year}
  {2021}{\natexlab{b}})}\BibitemShut {NoStop}%
\bibitem [{\citenamefont {{Shaymatov}}(2024)}]{Shaymatov24PRD.110d4042S}%
  \BibitemOpen
  \bibfield  {author} {\bibinfo {author} {\bibfnamefont {S.}~\bibnamefont
  {{Shaymatov}}},\ }\href {\doibase 10.1103/PhysRevD.110.044042} {\bibfield
  {journal} {\bibinfo  {journal} {Phys. Rev. D}\ }\textbf {\bibinfo {volume}
  {110}},\ \bibinfo {eid} {044042} (\bibinfo {year} {2024})},\ \Eprint
  {http://arxiv.org/abs/2402.02471} {arXiv:2402.02471 [gr-qc]} \BibitemShut
  {NoStop}%
\bibitem [{\citenamefont {{Piotrovich}}\ \emph {et~al.}(2010)\citenamefont
  {{Piotrovich}}, \citenamefont {{Silant'ev}}, \citenamefont {{Gnedin}},\ and\
  \citenamefont {{Natsvlishvili}}}]{Piotrovich10}%
  \BibitemOpen
  \bibfield  {author} {\bibinfo {author} {\bibfnamefont {M.~Y.}\ \bibnamefont
  {{Piotrovich}}}, \bibinfo {author} {\bibfnamefont {N.~A.}\ \bibnamefont
  {{Silant'ev}}}, \bibinfo {author} {\bibfnamefont {Y.~N.}\ \bibnamefont
  {{Gnedin}}}, \ and\ \bibinfo {author} {\bibfnamefont {T.~M.}\ \bibnamefont
  {{Natsvlishvili}}},\ }\href@noop {} {\bibfield  {journal} {\bibinfo
  {journal} {ArXiv e-prints}\ } (\bibinfo {year} {2010})},\ \Eprint
  {http://arxiv.org/abs/1002.4948} {arXiv:1002.4948 [astro-ph.CO]} \BibitemShut
  {NoStop}%
\bibitem [{\citenamefont {{Eatough}}\ and\ \citenamefont
  {et~al.}(2013)}]{Eatough13}%
  \BibitemOpen
  \bibfield  {author} {\bibinfo {author} {\bibfnamefont {R.~P.}\ \bibnamefont
  {{Eatough}}}\ and\ \bibinfo {author} {\bibnamefont {et~al.}},\ }\href
  {\doibase 10.1038/nature12499} {\bibfield  {journal} {\bibinfo  {journal}
  {Nature}\ }\textbf {\bibinfo {volume} {501}},\ \bibinfo {pages} {391}
  (\bibinfo {year} {2013})},\ \Eprint {http://arxiv.org/abs/1308.3147}
  {arXiv:1308.3147 [astro-ph.GA]} \BibitemShut {NoStop}%
\bibitem [{\citenamefont {{Shannon}}\ and\ \citenamefont
  {{Johnston}}(2013)}]{Shannon13}%
  \BibitemOpen
  \bibfield  {author} {\bibinfo {author} {\bibfnamefont {R.~M.}\ \bibnamefont
  {{Shannon}}}\ and\ \bibinfo {author} {\bibfnamefont {S.}~\bibnamefont
  {{Johnston}}},\ }\href {\doibase 10.1093/mnrasl/slt088} {\bibfield  {journal}
  {\bibinfo  {journal} {Mon. Not. R. Astron. Soc.}\ }\textbf {\bibinfo {volume}
  {435}},\ \bibinfo {pages} {L29} (\bibinfo {year} {2013})},\ \Eprint
  {http://arxiv.org/abs/1305.3036} {arXiv:1305.3036 [astro-ph.HE]} \BibitemShut
  {NoStop}%
\bibitem [{\citenamefont {{Baczko}}\ \emph {et~al.}(2016)\citenamefont
  {{Baczko}}, \citenamefont {{Schulz}},\ and\ \citenamefont {{et
  al.}}}]{Baczko16}%
  \BibitemOpen
  \bibfield  {author} {\bibinfo {author} {\bibfnamefont {A.-K.}\ \bibnamefont
  {{Baczko}}}, \bibinfo {author} {\bibfnamefont {R.}~\bibnamefont {{Schulz}}},
  \ and\ \bibinfo {author} {\bibnamefont {{et al.}}},\ }\href {\doibase
  10.1051/0004-6361/201527951} {\bibfield  {journal} {\bibinfo  {journal}
  {Astron. Astrophys.}\ }\textbf {\bibinfo {volume} {593}},\ \bibinfo {eid}
  {A47} (\bibinfo {year} {2016})},\ \Eprint {http://arxiv.org/abs/1605.07100}
  {arXiv:1605.07100} \BibitemShut {NoStop}%
\bibitem [{\citenamefont {{Dallilar}}\ and\ \citenamefont
  {et~al.}(2017)}]{Dallilar2018}%
  \BibitemOpen
  \bibfield  {author} {\bibinfo {author} {\bibfnamefont {Y.}~\bibnamefont
  {{Dallilar}}}\ and\ \bibinfo {author} {\bibnamefont {et~al.}},\ }\href
  {\doibase 10.1126/science.aan0249} {\bibfield  {journal} {\bibinfo  {journal}
  {Science}\ }\textbf {\bibinfo {volume} {358}},\ \bibinfo {pages} {1299}
  (\bibinfo {year} {2017})}\BibitemShut {NoStop}%
\bibitem [{\citenamefont {Akiyama}\ \emph
  {et~al.}(2021{\natexlab{a}})\citenamefont {Akiyama} \emph
  {et~al.}}]{EventHorizonTelescope:2021srq}%
  \BibitemOpen
  \bibfield  {author} {\bibinfo {author} {\bibfnamefont {K.}~\bibnamefont
  {Akiyama}} \emph {et~al.} (\bibinfo {collaboration} {Event Horizon
  Telescope}),\ }\href {\doibase 10.3847/2041-8213/abe4de} {\bibfield
  {journal} {\bibinfo  {journal} {Astrophys. J. Lett.}\ }\textbf {\bibinfo
  {volume} {910}},\ \bibinfo {pages} {L13} (\bibinfo {year}
  {2021}{\natexlab{a}})},\ \Eprint {http://arxiv.org/abs/2105.01173}
  {arXiv:2105.01173 [astro-ph.HE]} \BibitemShut {NoStop}%
\bibitem [{\citenamefont {Akiyama}\ \emph
  {et~al.}(2021{\natexlab{b}})\citenamefont {Akiyama} \emph
  {et~al.}}]{EventHorizonTelescope:2021bee}%
  \BibitemOpen
  \bibfield  {author} {\bibinfo {author} {\bibfnamefont {K.}~\bibnamefont
  {Akiyama}} \emph {et~al.} (\bibinfo {collaboration} {Event Horizon
  Telescope}),\ }\href {\doibase 10.3847/2041-8213/abe71d} {\bibfield
  {journal} {\bibinfo  {journal} {Astrophys. J. Lett.}\ }\textbf {\bibinfo
  {volume} {910}},\ \bibinfo {pages} {L12} (\bibinfo {year}
  {2021}{\natexlab{b}})},\ \Eprint {http://arxiv.org/abs/2105.01169}
  {arXiv:2105.01169 [astro-ph.HE]} \BibitemShut {NoStop}%
\bibitem [{\citenamefont {{Aliev}}\ and\ \citenamefont
  {{Gal'tsov}}(1989)}]{Aliev89}%
  \BibitemOpen
  \bibfield  {author} {\bibinfo {author} {\bibfnamefont {A.~N.}\ \bibnamefont
  {{Aliev}}}\ and\ \bibinfo {author} {\bibfnamefont {D.~V.}\ \bibnamefont
  {{Gal'tsov}}},\ }\href {\doibase 10.1070/PU1989v032n01ABEH002677} {\bibfield
  {journal} {\bibinfo  {journal} {Soviet Physics Uspekhi}\ }\textbf {\bibinfo
  {volume} {32}},\ \bibinfo {pages} {75} (\bibinfo {year} {1989})}\BibitemShut
  {NoStop}%
\bibitem [{\citenamefont {{Ernst}}(1976)}]{Ernst76}%
  \BibitemOpen
  \bibfield  {author} {\bibinfo {author} {\bibfnamefont {F.~J.}\ \bibnamefont
  {{Ernst}}},\ }\href {\doibase 10.1063/1.522781} {\bibfield  {journal}
  {\bibinfo  {journal} {Journal of Mathematical Physics}\ }\textbf {\bibinfo
  {volume} {17}},\ \bibinfo {pages} {54} (\bibinfo {year} {1976})}\BibitemShut
  {NoStop}%
\bibitem [{\citenamefont {{Gibbons}}\ \emph {et~al.}(2013)\citenamefont
  {{Gibbons}}, \citenamefont {{Mujtaba}},\ and\ \citenamefont
  {{Pope}}}]{Gibbons13}%
  \BibitemOpen
  \bibfield  {author} {\bibinfo {author} {\bibfnamefont {G.~W.}\ \bibnamefont
  {{Gibbons}}}, \bibinfo {author} {\bibfnamefont {A.~H.}\ \bibnamefont
  {{Mujtaba}}}, \ and\ \bibinfo {author} {\bibfnamefont {C.~N.}\ \bibnamefont
  {{Pope}}},\ }\href {\doibase 10.1088/0264-9381/30/12/125008} {\bibfield
  {journal} {\bibinfo  {journal} {Class. Quantum Grav.}\ }\textbf {\bibinfo
  {volume} {30}},\ \bibinfo {eid} {125008} (\bibinfo {year}
  {2013})}\BibitemShut {NoStop}%
\bibitem [{\citenamefont {Ernst}\ and\ \citenamefont {Wild}(1976)}]{Ernst76wz}%
  \BibitemOpen
  \bibfield  {author} {\bibinfo {author} {\bibfnamefont {F.}~\bibnamefont
  {Ernst}}\ and\ \bibinfo {author} {\bibfnamefont {W.}~\bibnamefont {Wild}},\
  }\href {\doibase 10.1063/1.522875} {\bibfield  {journal} {\bibinfo  {journal}
  {Journal of Mathematical Physics}\ }\textbf {\bibinfo {volume} {17}},\
  \bibinfo {pages} {182} (\bibinfo {year} {1976})},\ \Eprint
  {http://arxiv.org/abs/https://doi.org/10.1063/1.522875}
  {https://doi.org/10.1063/1.522875} \BibitemShut {NoStop}%
\bibitem [{\citenamefont {Aliev}\ and\ \citenamefont
  {Galtsov}(1989)}]{Aliev89wz}%
  \BibitemOpen
  \bibfield  {author} {\bibinfo {author} {\bibfnamefont {A.~N.}\ \bibnamefont
  {Aliev}}\ and\ \bibinfo {author} {\bibfnamefont {D.~V.}\ \bibnamefont
  {Galtsov}},\ }\href {\doibase 10.1007/BF00643854} {\bibfield  {journal}
  {\bibinfo  {journal} {Astrophys. Space Sci.}\ }\textbf {\bibinfo {volume}
  {155}},\ \bibinfo {pages} {181} (\bibinfo {year} {1989})}\BibitemShut
  {NoStop}%
\bibitem [{\citenamefont {{Garc{\'\i}a D{\'\i}az}}(1985)}]{Garcia85wz}%
  \BibitemOpen
  \bibfield  {author} {\bibinfo {author} {\bibfnamefont {A.}~\bibnamefont
  {{Garc{\'\i}a D{\'\i}az}}},\ }\href {\doibase 10.1063/1.526777} {\bibfield
  {journal} {\bibinfo  {journal} {Journal of Mathematical Physics}\ }\textbf
  {\bibinfo {volume} {26}},\ \bibinfo {pages} {155} (\bibinfo {year}
  {1985})}\BibitemShut {NoStop}%
\bibitem [{\citenamefont {Gibbons}\ \emph {et~al.}(2014)\citenamefont
  {Gibbons}, \citenamefont {Pang},\ and\ \citenamefont {Pope}}]{Gibbons14wz}%
  \BibitemOpen
  \bibfield  {author} {\bibinfo {author} {\bibfnamefont {G.~W.}\ \bibnamefont
  {Gibbons}}, \bibinfo {author} {\bibfnamefont {Y.}~\bibnamefont {Pang}}, \
  and\ \bibinfo {author} {\bibfnamefont {C.~N.}\ \bibnamefont {Pope}},\ }\href
  {\doibase 10.1103/PhysRevD.89.044029} {\bibfield  {journal} {\bibinfo
  {journal} {Phys. Rev. D}\ }\textbf {\bibinfo {volume} {89}},\ \bibinfo
  {pages} {044029} (\bibinfo {year} {2014})}\BibitemShut {NoStop}%
\bibitem [{\citenamefont {Astorino}\ \emph {et~al.}(2016)\citenamefont
  {Astorino}, \citenamefont {Comp\`ere}, \citenamefont {Oliveri},\ and\
  \citenamefont {Vandevoorde}}]{Astorino16wz}%
  \BibitemOpen
  \bibfield  {author} {\bibinfo {author} {\bibfnamefont {M.}~\bibnamefont
  {Astorino}}, \bibinfo {author} {\bibfnamefont {G.}~\bibnamefont {Comp\`ere}},
  \bibinfo {author} {\bibfnamefont {R.}~\bibnamefont {Oliveri}}, \ and\
  \bibinfo {author} {\bibfnamefont {N.}~\bibnamefont {Vandevoorde}},\ }\href
  {\doibase 10.1103/PhysRevD.94.024019} {\bibfield  {journal} {\bibinfo
  {journal} {Phys. Rev. D}\ }\textbf {\bibinfo {volume} {94}},\ \bibinfo
  {pages} {024019} (\bibinfo {year} {2016})}\BibitemShut {NoStop}%
\bibitem [{\citenamefont {{Konoplya}}\ and\ \citenamefont
  {{Fontana}}(2008)}]{Konoplya08a}%
  \BibitemOpen
  \bibfield  {author} {\bibinfo {author} {\bibfnamefont {R.~A.}\ \bibnamefont
  {{Konoplya}}}\ and\ \bibinfo {author} {\bibfnamefont {R.~D.~B.}\ \bibnamefont
  {{Fontana}}},\ }\href {\doibase 10.1016/j.physletb.2007.10.065} {\bibfield
  {journal} {\bibinfo  {journal} {Phys. Lett. B}\ }\textbf {\bibinfo {volume}
  {659}},\ \bibinfo {pages} {375} (\bibinfo {year} {2008})},\ \Eprint
  {http://arxiv.org/abs/0707.1156} {arXiv:0707.1156 [hep-th]} \BibitemShut
  {NoStop}%
\bibitem [{\citenamefont {{Konoplya}}(2008)}]{Konoplya08b}%
  \BibitemOpen
  \bibfield  {author} {\bibinfo {author} {\bibfnamefont {R.~A.}\ \bibnamefont
  {{Konoplya}}},\ }\href {\doibase 10.1016/j.physletb.2008.07.079} {\bibfield
  {journal} {\bibinfo  {journal} {Phys. Lett. B}\ }\textbf {\bibinfo {volume}
  {666}},\ \bibinfo {pages} {283} (\bibinfo {year} {2008})},\ \Eprint
  {http://arxiv.org/abs/0801.0846} {arXiv:0801.0846 [hep-th]} \BibitemShut
  {NoStop}%
\bibitem [{\citenamefont {{Shaymatov}}\ \emph
  {et~al.}(2021{\natexlab{c}})\citenamefont {{Shaymatov}}, \citenamefont
  {{Narzilloev}}, \citenamefont {{Abdujabbarov}},\ and\ \citenamefont
  {{Bambi}}}]{Shaymatov21c}%
  \BibitemOpen
  \bibfield  {author} {\bibinfo {author} {\bibfnamefont {S.}~\bibnamefont
  {{Shaymatov}}}, \bibinfo {author} {\bibfnamefont {B.}~\bibnamefont
  {{Narzilloev}}}, \bibinfo {author} {\bibfnamefont {A.}~\bibnamefont
  {{Abdujabbarov}}}, \ and\ \bibinfo {author} {\bibfnamefont {C.}~\bibnamefont
  {{Bambi}}},\ }\href {\doibase 10.1103/PhysRevD.103.124066} {\bibfield
  {journal} {\bibinfo  {journal} {Phys. Rev. D}\ }\textbf {\bibinfo {volume}
  {103}},\ \bibinfo {eid} {124066} (\bibinfo {year} {2021}{\natexlab{c}})},\
  \Eprint {http://arxiv.org/abs/2105.00342} {arXiv:2105.00342 [gr-qc]}
  \BibitemShut {NoStop}%
\bibitem [{\citenamefont {{Shaymatov}}\ \emph
  {et~al.}(2022{\natexlab{a}})\citenamefont {{Shaymatov}}, \citenamefont
  {{Sheoran}}, \citenamefont {{Becerril}}, \citenamefont {{Nucamendi}},\ and\
  \citenamefont {{Ahmedov}}}]{Shaymatov22PhRvD.106b4039S}%
  \BibitemOpen
  \bibfield  {author} {\bibinfo {author} {\bibfnamefont {S.}~\bibnamefont
  {{Shaymatov}}}, \bibinfo {author} {\bibfnamefont {P.}~\bibnamefont
  {{Sheoran}}}, \bibinfo {author} {\bibfnamefont {R.}~\bibnamefont
  {{Becerril}}}, \bibinfo {author} {\bibfnamefont {U.}~\bibnamefont
  {{Nucamendi}}}, \ and\ \bibinfo {author} {\bibfnamefont {B.}~\bibnamefont
  {{Ahmedov}}},\ }\href {\doibase 10.1103/PhysRevD.106.024039} {\bibfield
  {journal} {\bibinfo  {journal} {Phys. Rev. D}\ }\textbf {\bibinfo {volume}
  {106}},\ \bibinfo {eid} {024039} (\bibinfo {year}
  {2022}{\natexlab{a}})}\BibitemShut {NoStop}%
\bibitem [{\citenamefont {Mirkhaydarov}\ \emph {et~al.}(2026)\citenamefont
  {Mirkhaydarov}, \citenamefont {Xamidov}, \citenamefont {Sheoran},
  \citenamefont {Shaymatov},\ and\ \citenamefont
  {Nandan}}]{Mirkhaydarov2026MPP}%
  \BibitemOpen
  \bibfield  {author} {\bibinfo {author} {\bibfnamefont {M.}~\bibnamefont
  {Mirkhaydarov}}, \bibinfo {author} {\bibfnamefont {T.}~\bibnamefont
  {Xamidov}}, \bibinfo {author} {\bibfnamefont {P.}~\bibnamefont {Sheoran}},
  \bibinfo {author} {\bibfnamefont {S.}~\bibnamefont {Shaymatov}}, \ and\
  \bibinfo {author} {\bibfnamefont {H.}~\bibnamefont {Nandan}},\ }\href
  {https://arxiv.org/abs/2601.09919} {\enquote {\bibinfo {title} {Non-monotonic
  enhancement of the magnetic penrose process in kerr-bertotti-robinson
  spacetime and its implication for electron acceleration},}\ } (\bibinfo
  {year} {2026}),\ \Eprint {http://arxiv.org/abs/2601.09919} {arXiv:2601.09919
  [gr-qc]} \BibitemShut {NoStop}%
\bibitem [{\citenamefont {{Shaymatov}}\ and\ \citenamefont
  {{Ahmedov}}(2023)}]{Shaymatov23GRG}%
  \BibitemOpen
  \bibfield  {author} {\bibinfo {author} {\bibfnamefont {S.}~\bibnamefont
  {{Shaymatov}}}\ and\ \bibinfo {author} {\bibfnamefont {B.}~\bibnamefont
  {{Ahmedov}}},\ }\href {\doibase 10.1007/s10714-023-03082-y} {\bibfield
  {journal} {\bibinfo  {journal} {Gen. Relativ. Gravit.}\ }\textbf {\bibinfo
  {volume} {55}},\ \bibinfo {eid} {36} (\bibinfo {year} {2023})},\ \Eprint
  {http://arxiv.org/abs/2301.08569} {arXiv:2301.08569 [gr-qc]} \BibitemShut
  {NoStop}%
\bibitem [{\citenamefont {{Shaymatov}}\ \emph
  {et~al.}(2022{\natexlab{b}})\citenamefont {{Shaymatov}}, \citenamefont
  {{Jamil}}, \citenamefont {{Jusufi}},\ and\ \citenamefont
  {{Bamba}}}]{2022EPJC...82..636S}%
  \BibitemOpen
  \bibfield  {author} {\bibinfo {author} {\bibfnamefont {S.}~\bibnamefont
  {{Shaymatov}}}, \bibinfo {author} {\bibfnamefont {M.}~\bibnamefont
  {{Jamil}}}, \bibinfo {author} {\bibfnamefont {K.}~\bibnamefont {{Jusufi}}}, \
  and\ \bibinfo {author} {\bibfnamefont {K.}~\bibnamefont {{Bamba}}},\ }\href
  {\doibase 10.1140/epjc/s10052-022-10560-1} {\bibfield  {journal} {\bibinfo
  {journal} {Eur. Phys. J. C}\ }\textbf {\bibinfo {volume} {82}},\ \bibinfo
  {eid} {636} (\bibinfo {year} {2022}{\natexlab{b}})},\ \Eprint
  {http://arxiv.org/abs/2205.00270} {arXiv:2205.00270 [gr-qc]} \BibitemShut
  {NoStop}%
\bibitem [{\citenamefont {Alonso-Bardaji}\ \emph
  {et~al.}(2022{\natexlab{a}})\citenamefont {Alonso-Bardaji}, \citenamefont
  {Brizuela},\ and\ \citenamefont {Vera}}]{Alonso-Bardaji:2021yls}%
  \BibitemOpen
  \bibfield  {author} {\bibinfo {author} {\bibfnamefont {A.}~\bibnamefont
  {Alonso-Bardaji}}, \bibinfo {author} {\bibfnamefont {D.}~\bibnamefont
  {Brizuela}}, \ and\ \bibinfo {author} {\bibfnamefont {R.}~\bibnamefont
  {Vera}},\ }\href {\doibase 10.1016/j.physletb.2022.137075} {\bibfield
  {journal} {\bibinfo  {journal} {Phys. Lett. B}\ }\textbf {\bibinfo {volume}
  {829}},\ \bibinfo {pages} {137075} (\bibinfo {year} {2022}{\natexlab{a}})},\
  \Eprint {http://arxiv.org/abs/2112.12110} {arXiv:2112.12110 [gr-qc]}
  \BibitemShut {NoStop}%
\bibitem [{\citenamefont {Alonso-Bardaji}\ \emph
  {et~al.}(2022{\natexlab{b}})\citenamefont {Alonso-Bardaji}, \citenamefont
  {Brizuela},\ and\ \citenamefont {Vera}}]{Alonso-Bardaji:2022ear}%
  \BibitemOpen
  \bibfield  {author} {\bibinfo {author} {\bibfnamefont {A.}~\bibnamefont
  {Alonso-Bardaji}}, \bibinfo {author} {\bibfnamefont {D.}~\bibnamefont
  {Brizuela}}, \ and\ \bibinfo {author} {\bibfnamefont {R.}~\bibnamefont
  {Vera}},\ }\href {\doibase 10.1103/PhysRevD.106.024035} {\bibfield  {journal}
  {\bibinfo  {journal} {Phys. Rev. D}\ }\textbf {\bibinfo {volume} {106}},\
  \bibinfo {pages} {024035} (\bibinfo {year} {2022}{\natexlab{b}})},\ \Eprint
  {http://arxiv.org/abs/2205.02098} {arXiv:2205.02098 [gr-qc]} \BibitemShut
  {NoStop}%
\bibitem [{\citenamefont {Levin}\ and\ \citenamefont
  {Perez-Giz}(2009)}]{Levin:2008yp}%
  \BibitemOpen
  \bibfield  {author} {\bibinfo {author} {\bibfnamefont {J.}~\bibnamefont
  {Levin}}\ and\ \bibinfo {author} {\bibfnamefont {G.}~\bibnamefont
  {Perez-Giz}},\ }\href {\doibase 10.1103/PhysRevD.79.124013} {\bibfield
  {journal} {\bibinfo  {journal} {Phys. Rev. D}\ }\textbf {\bibinfo {volume}
  {79}},\ \bibinfo {pages} {124013} (\bibinfo {year} {2009})},\ \Eprint
  {http://arxiv.org/abs/0811.3814} {arXiv:0811.3814 [gr-qc]} \BibitemShut
  {NoStop}%
\bibitem [{\citenamefont {Chandrasekhar}(1985)}]{Chandrasekhar:1985kt}%
  \BibitemOpen
  \bibfield  {author} {\bibinfo {author} {\bibfnamefont {S.}~\bibnamefont
  {Chandrasekhar}},\ }\href@noop {} {\emph {\bibinfo {title} {{The mathematical
  theory of black holes}}}}\ (\bibinfo {year} {1985})\BibitemShut {NoStop}%
\bibitem [{\citenamefont {Hughes}(2000)}]{Hughes:1999bq}%
  \BibitemOpen
  \bibfield  {author} {\bibinfo {author} {\bibfnamefont {S.~A.}\ \bibnamefont
  {Hughes}},\ }\href {\doibase 10.1103/PhysRevD.65.069902} {\bibfield
  {journal} {\bibinfo  {journal} {Phys. Rev. D}\ }\textbf {\bibinfo {volume}
  {61}},\ \bibinfo {pages} {084004} (\bibinfo {year} {2000})},\ \bibinfo {note}
  {[Erratum: Phys.Rev.D 63, 049902 (2001), Erratum: Phys.Rev.D 65, 069902
  (2002), Erratum: Phys.Rev.D 67, 089901 (2003), Erratum: Phys.Rev.D 78, 109902
  (2008), Erratum: Phys.Rev.D 90, 109904 (2014)]},\ \Eprint
  {http://arxiv.org/abs/gr-qc/9910091} {arXiv:gr-qc/9910091} \BibitemShut
  {NoStop}%
\bibitem [{\citenamefont {Barack}\ and\ \citenamefont
  {Cutler}(2004)}]{Barack:2003fp}%
  \BibitemOpen
  \bibfield  {author} {\bibinfo {author} {\bibfnamefont {L.}~\bibnamefont
  {Barack}}\ and\ \bibinfo {author} {\bibfnamefont {C.}~\bibnamefont
  {Cutler}},\ }\href {\doibase 10.1103/PhysRevD.69.082005} {\bibfield
  {journal} {\bibinfo  {journal} {Phys. Rev. D}\ }\textbf {\bibinfo {volume}
  {69}},\ \bibinfo {pages} {082005} (\bibinfo {year} {2004})},\ \Eprint
  {http://arxiv.org/abs/gr-qc/0310125} {arXiv:gr-qc/0310125} \BibitemShut
  {NoStop}%
\bibitem [{\citenamefont {Isoyama}\ \emph {et~al.}(2022)\citenamefont
  {Isoyama}, \citenamefont {Fujita}, \citenamefont {Chua}, \citenamefont
  {Nakano}, \citenamefont {Pound},\ and\ \citenamefont
  {Sago}}]{Isoyama:2021jjd}%
  \BibitemOpen
  \bibfield  {author} {\bibinfo {author} {\bibfnamefont {S.}~\bibnamefont
  {Isoyama}}, \bibinfo {author} {\bibfnamefont {R.}~\bibnamefont {Fujita}},
  \bibinfo {author} {\bibfnamefont {A.~J.~K.}\ \bibnamefont {Chua}}, \bibinfo
  {author} {\bibfnamefont {H.}~\bibnamefont {Nakano}}, \bibinfo {author}
  {\bibfnamefont {A.}~\bibnamefont {Pound}}, \ and\ \bibinfo {author}
  {\bibfnamefont {N.}~\bibnamefont {Sago}},\ }\href {\doibase
  10.1103/PhysRevLett.128.231101} {\bibfield  {journal} {\bibinfo  {journal}
  {Phys. Rev. Lett.}\ }\textbf {\bibinfo {volume} {128}},\ \bibinfo {pages}
  {231101} (\bibinfo {year} {2022})},\ \Eprint
  {http://arxiv.org/abs/2111.05288} {arXiv:2111.05288 [gr-qc]} \BibitemShut
  {NoStop}%
\bibitem [{\citenamefont {Babak}\ \emph {et~al.}(2007)\citenamefont {Babak},
  \citenamefont {Fang}, \citenamefont {Gair}, \citenamefont {Glampedakis},\
  and\ \citenamefont {Hughes}}]{Babak:2006uv}%
  \BibitemOpen
  \bibfield  {author} {\bibinfo {author} {\bibfnamefont {S.}~\bibnamefont
  {Babak}}, \bibinfo {author} {\bibfnamefont {H.}~\bibnamefont {Fang}},
  \bibinfo {author} {\bibfnamefont {J.~R.}\ \bibnamefont {Gair}}, \bibinfo
  {author} {\bibfnamefont {K.}~\bibnamefont {Glampedakis}}, \ and\ \bibinfo
  {author} {\bibfnamefont {S.~A.}\ \bibnamefont {Hughes}},\ }\href {\doibase
  10.1103/PhysRevD.75.024005} {\bibfield  {journal} {\bibinfo  {journal} {Phys.
  Rev. D}\ }\textbf {\bibinfo {volume} {75}},\ \bibinfo {pages} {024005}
  (\bibinfo {year} {2007})},\ \bibinfo {note} {[Erratum: Phys.Rev.D 77, 04990
  (2008)]},\ \Eprint {http://arxiv.org/abs/gr-qc/0607007} {arXiv:gr-qc/0607007}
  \BibitemShut {NoStop}%
\bibitem [{\citenamefont {Hughes}(2001)}]{Hughes:2000ssa}%
  \BibitemOpen
  \bibfield  {author} {\bibinfo {author} {\bibfnamefont {S.~A.}\ \bibnamefont
  {Hughes}},\ }\href {\doibase 10.1088/0264-9381/18/19/314} {\bibfield
  {journal} {\bibinfo  {journal} {Class. Quant. Grav.}\ }\textbf {\bibinfo
  {volume} {18}},\ \bibinfo {pages} {4067} (\bibinfo {year} {2001})},\ \Eprint
  {http://arxiv.org/abs/gr-qc/0008058} {arXiv:gr-qc/0008058} \BibitemShut
  {NoStop}%
\bibitem [{\citenamefont {Thorne}(1980)}]{Thorne:1980ru}%
  \BibitemOpen
  \bibfield  {author} {\bibinfo {author} {\bibfnamefont {K.~S.}\ \bibnamefont
  {Thorne}},\ }\href {\doibase 10.1103/RevModPhys.52.299} {\bibfield  {journal}
  {\bibinfo  {journal} {Rev. Mod. Phys.}\ }\textbf {\bibinfo {volume} {52}},\
  \bibinfo {pages} {299} (\bibinfo {year} {1980})}\BibitemShut {NoStop}%
\bibitem [{\citenamefont {Gair}\ \emph {et~al.}(2017)\citenamefont {Gair},
  \citenamefont {Babak}, \citenamefont {Sesana}, \citenamefont {Amaro-Seoane},
  \citenamefont {Barausse}, \citenamefont {Berry}, \citenamefont {Berti},\ and\
  \citenamefont {Sopuerta}}]{Gair:2017ynp}%
  \BibitemOpen
  \bibfield  {author} {\bibinfo {author} {\bibfnamefont {J.~R.}\ \bibnamefont
  {Gair}}, \bibinfo {author} {\bibfnamefont {S.}~\bibnamefont {Babak}},
  \bibinfo {author} {\bibfnamefont {A.}~\bibnamefont {Sesana}}, \bibinfo
  {author} {\bibfnamefont {P.}~\bibnamefont {Amaro-Seoane}}, \bibinfo {author}
  {\bibfnamefont {E.}~\bibnamefont {Barausse}}, \bibinfo {author}
  {\bibfnamefont {C.~P.~L.}\ \bibnamefont {Berry}}, \bibinfo {author}
  {\bibfnamefont {E.}~\bibnamefont {Berti}}, \ and\ \bibinfo {author}
  {\bibfnamefont {C.}~\bibnamefont {Sopuerta}},\ }\href {\doibase
  10.1088/1742-6596/840/1/012021} {\bibfield  {journal} {\bibinfo  {journal}
  {J. Phys. Conf. Ser.}\ }\textbf {\bibinfo {volume} {840}},\ \bibinfo {pages}
  {012021} (\bibinfo {year} {2017})},\ \Eprint
  {http://arxiv.org/abs/1704.00009} {arXiv:1704.00009 [astro-ph.GA]}
  \BibitemShut {NoStop}%
\bibitem [{\citenamefont {Babak}\ \emph {et~al.}(2017)\citenamefont {Babak},
  \citenamefont {Gair}, \citenamefont {Sesana}, \citenamefont {Barausse},
  \citenamefont {Sopuerta}, \citenamefont {Berry}, \citenamefont {Berti},
  \citenamefont {Amaro-Seoane}, \citenamefont {Petiteau},\ and\ \citenamefont
  {Klein}}]{Babak:2017tow}%
  \BibitemOpen
  \bibfield  {author} {\bibinfo {author} {\bibfnamefont {S.}~\bibnamefont
  {Babak}}, \bibinfo {author} {\bibfnamefont {J.}~\bibnamefont {Gair}},
  \bibinfo {author} {\bibfnamefont {A.}~\bibnamefont {Sesana}}, \bibinfo
  {author} {\bibfnamefont {E.}~\bibnamefont {Barausse}}, \bibinfo {author}
  {\bibfnamefont {C.~F.}\ \bibnamefont {Sopuerta}}, \bibinfo {author}
  {\bibfnamefont {C.~P.~L.}\ \bibnamefont {Berry}}, \bibinfo {author}
  {\bibfnamefont {E.}~\bibnamefont {Berti}}, \bibinfo {author} {\bibfnamefont
  {P.}~\bibnamefont {Amaro-Seoane}}, \bibinfo {author} {\bibfnamefont
  {A.}~\bibnamefont {Petiteau}}, \ and\ \bibinfo {author} {\bibfnamefont
  {A.}~\bibnamefont {Klein}},\ }\href {\doibase 10.1103/PhysRevD.95.103012}
  {\bibfield  {journal} {\bibinfo  {journal} {Phys. Rev. D}\ }\textbf {\bibinfo
  {volume} {95}},\ \bibinfo {pages} {103012} (\bibinfo {year} {2017})},\
  \Eprint {http://arxiv.org/abs/1703.09722} {arXiv:1703.09722 [gr-qc]}
  \BibitemShut {NoStop}%
\end{thebibliography}%

\end{document}